\documentclass[manuscript, screen]{acmart}
\AtBeginDocument{%
  }

\setcopyright{acmlicensed}
\copyrightyear{2025}
\acmYear{2025}
\acmDOI{XXXXXXX.XXXXXXX}

\acmJournal{TOMS}
\acmVolume{1}
\acmNumber{1}
\acmArticle{1}
\acmMonth{1}

\usepackage{subcaption}
\usepackage{tikz}
\usepackage{forest} 
\usepackage{xcolor}
\usepackage{subcaption}
\usepackage{booktabs,makecell}  
\usepackage{array}
\usepackage{algorithm}
\usepackage{algpseudocode}
\usepackage{pifont}
\usepackage{float}
\usetikzlibrary{arrows.meta}
\newcommand{\cmark}{\ding{51}}
\newcommand{\xmark}{\ding{55}}

\begin{document}

\title{Tree-Based Deep Learning for Ranking Symbolic Integration Algorithms}

\author{Rashid Barket}
\email{barketr@coventry.ac.uk}
\orcid{0000-0002-9104-4281} 
\author{Matthew England}
\email{matthew.england@coventry.ac.uk}
\orcid{0000-0001-5729-3420}
\affiliation{%
  \institution{Coventry University}
  \city{Coventry}
  \country{United Kingdom}
}

\author{J{\"u}rgen Gerhard}
\affiliation{%
  \institution{Maplesoft}
  \city{Waterloo}
  \country{Canada}
}
\email{jgerhard@maplesoft.com}

\begin{abstract}
Symbolic indefinite integration in Computer Algebra Systems such as Maple involves selecting the most effective algorithm from multiple available methods. Not all methods will succeed for a given problem, and when several do, the results, though mathematically equivalent, can differ greatly in presentation complexity. Traditionally, this choice has been made with minimal consideration of the problem instance, leading to inefficiencies.

We present a machine learning (ML) approach using tree-based deep learning models within a two-stage architecture: first identifying applicable methods for a given instance, then ranking them by predicted output complexity. Furthermore, we find representing mathematical expressions as tree structures significantly improves performance over sequence-based representations, and our two-stage framework outperforms alternative ML formulations.

Using a diverse dataset generated by six distinct data generators, our models achieve nearly 90\% accuracy in selecting the optimal method on a 70,000-example hold-out test set. On an independent out-of-distribution benchmark from Maple’s internal test suite, our tree transformer model maintains strong generalisation, outperforming Maple’s built-in selector and prior ML approaches.

These results highlight the critical role of data representation and problem framing in ML for symbolic computation, and we expect our methodology to generalise effectively to similar optimisation problems in mathematical software.
\end{abstract}


\renewcommand{\shortauthors}{Barket et al.}
\acmArticleType{Research}
\acmCodeLink{https://github.com/rbarket/Ranking_Symbolic_Int_Methods}
\acmDataLink{https://zenodo.org/records/16780106}
\acmContributions{RB, ME and JG all contributed to the study design and results analysis.  RB conducted the experiments and wrote the first draft of the paper, which ME and JG then assisted in editing.}

\maketitle

\section{Introduction}
\label{sec:intro}

Machine Learning (ML), particularly deep learning, has been applied in many domains, but its role in mathematical software is under-explored. In this context, ML can either seek to directly perform mathematical computations or to enhance algorithmic efficiency by optimising existing software. We are concerned specifically with Computer Algebra Systems (CASs), such as Maple.  While there has been some recent work on direct computation here \cite{kera2024_groebner}, most users choose these tools for the guaranteed exact results, often relying on these for scientific discovery or safety critical situations.  Thus there is little appetite for direct ML computation without complementary result validation.  Even an ML model with 99\% accuracy is not suitable for replacing a CAS algorithm if we insist on having exact results that are always correct.

\begin{figure}
    \centering
    \includegraphics[width=0.6\linewidth]{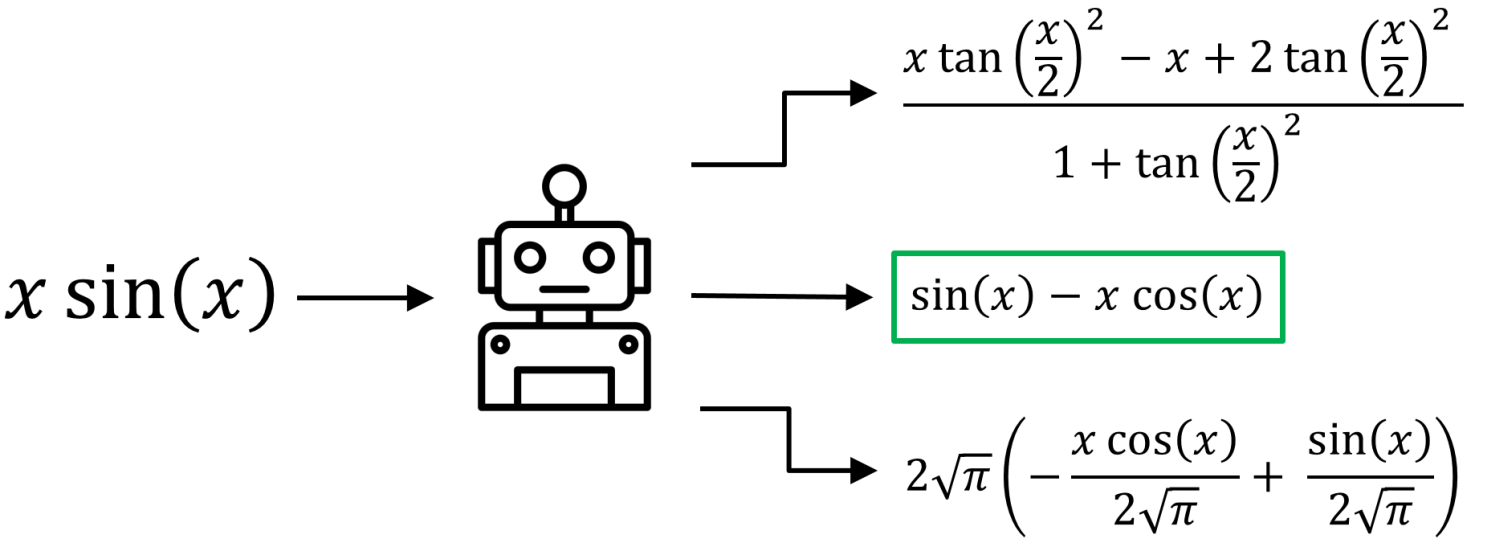}
    \caption{An example of three possible answers produced by different integration methods in Maple when integrating $x\sin(x)$. Although all are mathematically equivalent, we want ML to choose the method giving the ``simplest'' answer.}
    \label{fig:example_output}
    \Description{An example of integrating a mathematical expression in Maple and it can produce several different answers. All of them are correct, but the one that is the most intuitive to the user is the one we wish to output.}
\end{figure}

We focus instead on using ML to improve the performance of CASs without risking the mathematical correctness of their results.  Specifically, we consider Maple's symbolic integration user algorithm, which in turn has access to 13 integration ``\emph{methods}''.  For details on these methods see, for example, the online Maple help page titled ``\emph{
Integration Methods}''.  When such methods succeed their results are mathematically equivalent, but they can vary considerable in their presentation. A poor choice of method can lead to longer computation time and unnecessarily intricate expressions. A small example of these different presentations is shown in Figure \ref{fig:example_output} (see also Section \ref{sec:results/compare} later for some examples where the differences are more substantial and not bridged by mathematical simplification tools). Guided by our industrial partner Maplesoft, who produce the Maple CAS, we place our focus on the output representation.  However, we expect that all our work could be repurposed to focus on runtime instead if so desired.

Maple currently tries these integration methods in sequence until one succeeds for a problem, in some cases excluding methods based on simple input checks where this is thought to conclude a method's applicability in advance.  It has already been observed that ML techniques common in natural language processing, such as LSTMs and transformers, are able to perform well on datasets of mathematical expressions, which can be easily represented as a sequence of tokens. In particular, transformers showed promising results in the direct computation of integrals \cite{Lample2020}. It is thus natural to consider if these tools can be used to select the ``best'' of the available integration methods in Maple for specific problem instances.  

\subsection{Contributions}

We will present new state-of-the-art solutions for this task.  We demonstrate that there are significant differences in performance available depending on how we frame this problem for ML.  We have previously reported on the use of LSTMs to solve this as a classification problem (selecting the best method) \cite{Barket2024_TreeLSTM}, but we report here substantial improvements from the use of a two-stage architecture to first identify admissible methods and then rank them. 
We also explore here the effect of data representation: rather than simply representing a mathematical expression as a list of characters or tokens, they can be quite naturally represented as a tree structure, which encodes hierarchical information about the expression.  When comparing the sequence and tree variants of ML models (both LSTMs and transformers) we will see that the tree-based methods give improved performance. We expect that the lessons of this paper may apply to similar decisions in other CASs, or indeed other types of mathematical software.

\subsection{Plan of the Paper}

We continue in Section~\ref{sec:litreview} by reviewing the literature on ML for symbolic integration.  Then Section~\ref{sec:data} describes the different ways to represent data for an ML model and describes our dataset: the data generators we use have been discussed in prior work \cite{Barket2023_generation, Lample2020, Barket2024_Liouville}, but here we extend them to work with non-elementary functions as described in Appendix~\ref{sec:special_functions}.  The main paper continues in Section~\ref{sec:architecture}  where we explain the difference between the sequence and tree-based ML models we use, and then Section~\ref{sec:methodology} where we describe two different ways to frame the method selection problem (a third intermediate framing, regression, is included in Appendix~\ref{sec:regression}). Finally, all experimental results are presented in Section~\ref{sec:results}, before we conclude and discuss future work in Section~\ref{sec:conclusion}. 

\section{Literature Review}
\label{sec:litreview}

\subsection{ML for Direct Symbolic Integration}

Lample and Charton \cite{Lample2020} were the first to use ML to perform symbolic integration directly in 2020. They represent mathematical expressions as sequences and generated large synthetic datasets through three different data generation methods --- FWD, BWD and IBP --- which we describe later in Section~\ref{sec:data/exisiting}. They trained transformer models on these datasets, and reported that the transformers surpassed commercial CASs Mathematica and Maple in producing a correct answer within a reasonable time, although this evaluation conflated the different situations of timeout and incorrect output (there were cases were the commercial CASs timed-out, but only the transformers produced incorrect answers).  

Although a seminal work in the field, there are several drawbacks to the methodology that have been identified since.  First, the authors themselves reported that the models perform poorly on out-of-distribution data (e.g. training on FWD data and testing on BWD data).  Later, certain biases in their dataset were identified (see \cite{Davis2019_critique, Piotrowski2019, Barket2024_Liouville}) which both suggest a constrained distribution of problems and the risk of data leakage between training and testing.  Nonetheless, it was an impressive feat to train transformers to solve problem instances that CASs cannot, and spurred interest in the broader field of ML for mathematical computation.

Welleck et al. \cite{Welleck2022_Brittleness} examined the limitations of generative sequence models in symbolic integration. They found that while sequence models can achieve near-perfect accuracy on test sets, they have significant weaknesses in out-of-distribution (OOD) generalisation.  The authors developed a genetic algorithm, SAGGA, that identifies adversarial failures, to demonstrate that small perturbations to inputs on which the model is successful cannot be tackled, allowing for structured evaluation of robustness.  They also demonstrate an issues around composition:  models regularly fail to integrate the sum of two smaller expressions that it could integrate.  The study concludes that current deep learning approaches, which rely on maximum likelihood estimation, lack the structured reasoning necessary for symbolic mathematics, emphasising the need for architectures with stronger inductive biases and generalization capabilities.

\subsection{ML for Optimisation of Symbolic Integration}

Efforts have been put into ML optimisation of symbolic integration algorithms, as opposed than ML integrating directly. The present authors reported on preliminary work last year comparing LSTM and TreeLSTM models for our problem of selecting symbolic integration methods in Maple \cite{Barket2024_TreeLSTM}. The TreeLSTM used a tree version of the dataset, whereas the regular LSTM a sequence version and otherwise were trained under the exact same conditions (the only difference being a layer in the underlying model architecture). Once trained, we compare the models against each other and found the tree architecture offered significant advantages.  We recreate those experiments as part of the present paper $-$ here we use a significantly updated dataset (see e.g. Figure~\ref{fig:results/bin_cls}).  Although the TreeLSTM model was superior to Maple's existing approach on hold-out test data, in \cite{Barket2024_TreeLSTM} we found that the performance drop when moving to out-of-distribution data was significant enough not to conclude overall superiority.  The additional combination of ML techniques we present in the present paper will allow us to make that claim, as reported in Section~\ref{sec:results}.

The Python-based CAS, SymPy, has also seen work in ML optimisation of its symbolic integration algorithm. Sharma et al. \cite{Sharma2023_SIRD} introduce SIRD-2M, a large-scale dataset comprising 2 million (function, integration-rule) pairs, aimed at improving the interpretability and efficiency of deep learning models for symbolic integration. Unlike previous black-box approaches that predict integrals directly, their method models integration as a step-by-step rule prediction task, making the process more transparent and structured. They train a transformer model on SIRD-2M and integrate it into SymPy's \texttt{integral\_steps} function, leading to a $6\times$ reduction in search space and a $2.28\times$ speed-up over the default heuristic-based search. While their approach enhances interpretability and efficiency, it is limited by its reliance on predefined integration rules and does not support certain function types, such as hyperbolic trigonometric functions. While their model generalises beyond its training data, it still falls short of more advanced symbolic solvers that leverage algorithms like Risch-Norman integration. Maple has a similar step-by-step integrator (in the \texttt{Student} library); if the work done on SymPy could be replicated in a more sophisticated CAS such as Maple it may overcome this shortcoming.

\subsection{Tree-based ML for Mathematics}
\label{subsec:LRTree}

Representing mathematical information as a tree structure is not a new concept; however, other than our preliminary work \cite{Barket2024_TreeLSTM}, few have exploited the structure in neural networks. For example, Wang et al. \cite{Wang2021_mathembed} and Wankerl et al. \cite{Wanker2021_polynoms} have explored the use of tree embeddings in ML for mathematics. Wang et al. train a gated recurrent neural network \cite{Cho2014_GRU} to perform formula reconstruction and formula retrieval with great success. Wankerl et. al train a tree-recursive neural network and a TreeLSTM to predict properties of polynomials, such as equivalence, derivative, and permutation of variables. However, neither work has yet been applied to more recent ML architectures such as transformers.  
The theorem-proving community has also been aware of the benefit from representing formulae as trees. Blaauwbroek et al. \cite{Blaauwbroek2024_TheoremProving} discuss multiple ways that ML have been used in theorem proving, including how tree-recursive neural networks have been used for tasks such as premise selection. More recently, Kogkalidis et al. \cite{Kogkalidis2024_algebraicPE} proposed a new way to represent trees based on group theory and have applied this to the Agda proof assistant, where they represent dependently-typed programs as trees \cite{Kogkalidis2024_Agda}. 

\section{Data}
\label{sec:data}

\subsection{Representing Mathematical Expressions}
\label{sec:data/representation}

\begin{figure}[t]
    \centering    \includegraphics[scale=0.5]{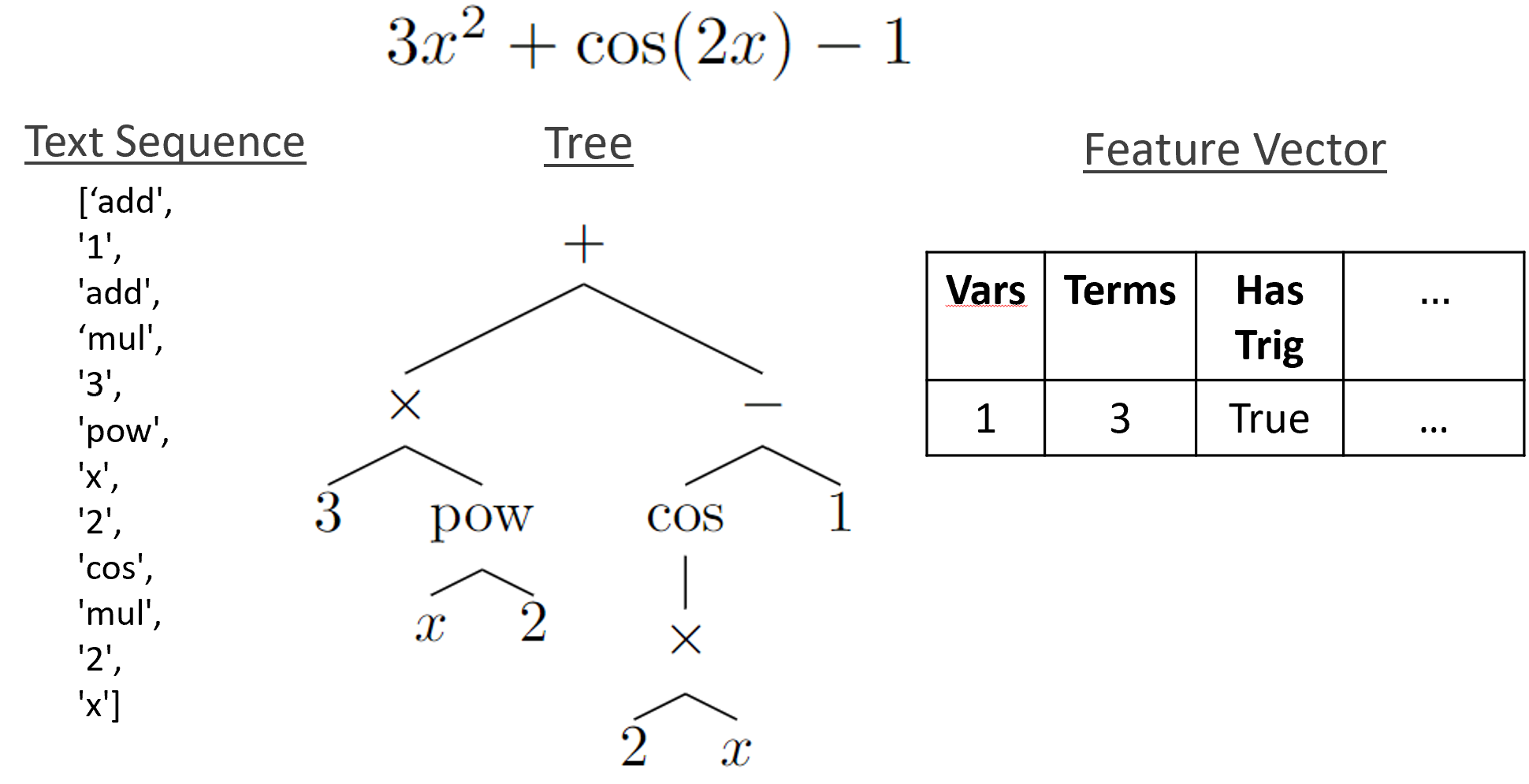}
    \caption{Different representations of the expression $3x^2~+~\cos(2x)~-1.$ Each would be suitable for a different type of ML model.}
    \Description{Possible representations of the expression $3x^2~+~\cos(2x)~-1.$}
    \label{fig:representations}
\end{figure}

To produce a model to do method selection, one must first be able to give training data to a model. There are various ways to represent mathematical expressions to an ML model, in turn allowing for experimentation with various types of ML models. Figure~\ref{fig:representations} shows three possible representations for mathematical expressions. 

The approach on the left is to use a sequence of tokens (numbers, variables, functions, and operators). A simple approach would be to do this directly from the common infix notation used in mathematical writing.  E.g. $1+\cos(x)$ becomes \texttt{"1,~+,~cos,~(,~x,~)"}.  However, ML literature has tended to use instead \textbf{prefix notation} (also known as Polish notation) for greater efficiency.  E.g. $1+\cos(x)$ becomes instead \texttt{"+,~1,~cos,~x"}.  In prefix notation we can avoid using brackets with no information lost. This allows for smaller models and reduced computational costs.  A sequential representation is suitable for modern deep learning methods that have emerged as powerful for natural language processing, such as LSTMs and transformers.

The approach on the right is to generate a vector of features to represent the expression.  For example, we might store the number of variables, the number of terms, and so on. This approach requires a domain expert to hand-craft each feature, but even then, there is no guarantee that the feature vector is of sufficient quality.  The feature vector tends not to encode all the raw information, which may hinder learning, although this can also guard against over-fitting.  A feature representation is suitable for traditional machine learning models such as support vector machines and decision trees.  These have the advantage of requiring far less data to accurately train and have been central to ML for CAS optimisation where data labelling is prohibitively expensive \cite{HEWBDP19}, \cite{Del2024_lessons}. Such models are also more open to explanation as to how they make their decisions \cite{Pickering2024_CAD}. 

The approach in the centre of the figure is to use the tree representation of an expression. This is a natural representation for a mathematical expression which captures the hierarchical and nested structure inherent, and yet surprisingly under-explored in the literature as described in Section~\ref{sec:litreview}. Here, the inner nodes of the tree are unary and binary operators or functions, and the leaves are variables and constants. The tree form of an expression can be used to train graph neural networks, and the tree variants of LSTMs and transformers.
 
\subsection{Generating Data}
\label{sec:data/generation}

Training a model to predict the optimal method for integration requires a diverse and well-structured dataset of integrable mathematical expressions. However, generating such expressions is non-trivial, and as described in Section~\ref{sec:litreview}, previous explored methods can lack sufficient variety, limiting the model’s ability to generalise beyond its training set. We describe our approach to generating data, summarising the work in our prior papers dedicated to this topic \cite{Barket2023_generation}, \cite{Barket2024_Liouville}.

\subsubsection{The generators of Lample and Charton (2020)}
\label{sec:data/exisiting}

The authors of \cite{Lample2020} were the first to describe data generation methods for indefinite integration. They presented three methods to produce (integrand, integral) pairs. 
\begin{itemize}
    \item \textbf{FWD:} Integrate a randomly generated expression $f$ with a CAS to get $F$, and add the pair ($f, F$) to the dataset.
    \item \textbf{BWD:} Differentiate a randomly generated expression $f$ to get $f'$ and add the pair ($f', f$) to the dataset.
    \item \textbf{IBP:} Given two expressions $f$ and $g$, calculate $f'$ and $g'$. If $\int f'g$ is known (i.e. exists already in our dataset) then the following holds (integration-by-parts):
    $\int fg' = fg - \int f'g$.
    Add the pair ($fg'$, $fg - \int f'g$) to the dataset. 
\end{itemize}

While this can generate a sufficiently large quantity of data, there exists a problem of insufficient variety of data. Ourselves and others have discussed these issues (\cite{Davis2019_critique, Piotrowski2019, Barket2023_generation}). We briefly overview the problems here:
\begin{itemize}
    \item There is a bias in the length of the expressions where a generator can produce long integrands and short integrals or vice-versa but not pairs of equal length.
    \item We can generated integrands which are too similar within the same dataset (i.e. differing only by their coefficients).
    \item These generators struggle to generate integrands with specific shape (see \cite{Barket2024_Liouville} for details). 
\end{itemize}

\subsubsection{Recent data generators from the present authors}

To address these concerns, we previously developed two new different data generators, taking inspiration from classical computer algebra work and in particular Liouville's theorem (see e.g. Section $12.4$ in \cite{Geddes1992}).

\begin{theorem}[Liouville's theorem] \label{thm:Liouville}
    Let $D$ be a differential field with constant field $K$ (whose algebraic closure is $L$). Suppose $f \in D$ and there exists $g$, elementary over $D$, such that $g'=f$. Then, $\exists \, v_0,\dots,v_m \in D, c_1,\dots,c_m \in L$ such that
    \begin{equation*}
        f=v_{0}' + \sum_{i=1}^{m} c_i\frac{v_{i}'}{v_i} \implies g = v_0 + \sum_{i=1}^{m} c_{i}\log(v_{i}).      
    \end{equation*}
\end{theorem}

This gives a general form to an elementary integral, if it exists. It led to our development of two new data generators.

\begin{itemize}
    \item \textbf{RISCH} \cite{Barket2023_generation}: produce an expression with symbolic coefficients, and go through the steps of the Risch algorithm \cite{Risch1969} to uncover the constraints on the symbolic coefficients needed for the expression to be integrable:  we then specify the coefficients to satisfy these and generate data.
    \item \textbf{LIOUVILLE} \cite{Barket2024_Liouville}: produce a general structured expression and follow the steps of the parallel Risch algorithm \cite{Norman1977_parallel} to integrate and produce a solution space for the symbolic coefficients. 
\end{itemize}

The correctness of the Risch algorithm is underpinned by Theorem~\ref{thm:Liouville}, while the parallel Risch algorithm is a heuristic approach inspired by the result.  We named the LIOUVILLE generator since the output takes the same format as $g$ there (i.e. an expression plus a sum of logarithms).  
We introduce one final data generator based on the substitution rule from calculus which, like IBP, requires an existing dataset.
\begin{itemize}
    \item \textbf{SUB} \cite{Barket2024_TreeLSTM}: Given $f$ and $g$, calculate $g'$ and let $u=g(x)$. If $\int f$ is already known, then the following holds (substitution rule): $\int f(g(x))g'(x) dx = \int f(u) du$. Add ($f(u), \int f(u)$) to the dataset.
\end{itemize}
We described in \cite{Barket2023_generation}, \cite{Barket2024_Liouville} how these new generators address the problems outlined above with the  original ones.

\subsubsection{Extending the data generators to non-elementary functions}

These six data generators above were all presented in prior literature where they were demonstrated to generate elementary functions (i.e. polynomials, rational functions, trigonometric functions, hyperbolic functions, exponential functions functions, and their inverses).  However, Maple also supports many non-elementary (special) functions which are in scope of our application.  In the present paper, we use a dataset containing both elementary and non-elementary functions.  We describe how we modified the existing data generators to produce non-elementary (integrand, integral) pairs in Appendix~\ref{sec:special_functions}.

\subsection{Data Labelling}
\label{sec:data/preparation}

Once we have data, we need to label it according to how the integration methods perform.  To do this, we need a metric to evaluate the method's performance on the data.  As described in the introduction, we focus on the output representation: an obvious metric is then the output length (or number of tokens). An alternative would be to try and characterise the complexity of those tokens:  Rich \cite{Rich2018_RUBI} defined their own metric based on a mix of size and the use of different types of functions (e.g. the integrand is elementary while the integral has a non-elementary function).
Maple stores all its expressions internally as directed acyclic graphs (DAGs), similar to expression trees but storing identical sub-expressions only once. Taking the DAG size as a metric is thus natural when working as Maple and will measure memory while being highly correlated to expression length. To measure the DAG size, we use a custom Maple function: 
\begin{verbatim}
    expr -> length(sprintf("%lm",expr))
\end{verbatim}
which is measuring the length of the serialised format of the DAG in Maple. We train ML models to optimise this.  We note is that while our metric is DAG size, the tree ML methods we use take the input data as an actual tree (i.e. with any identical sub-expressions duplicated). 

\begin{figure}[ht]
    \centering
    \includegraphics[width=0.8\linewidth]{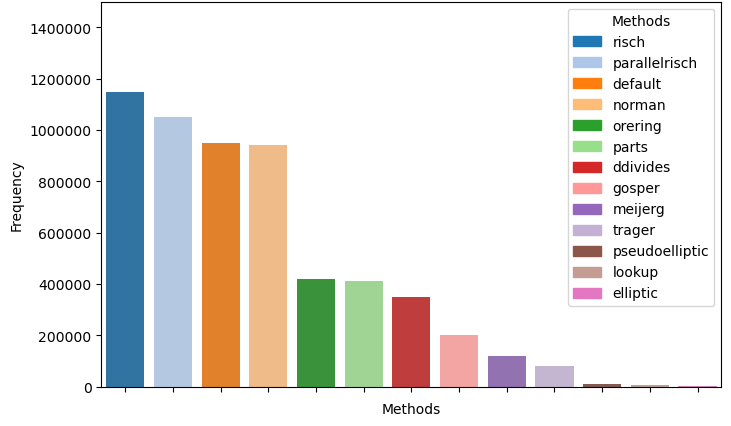}
    \caption{Frequency of each method being optimal over 1.5M examples from the six data generators. Note that more than one method may be optimal for a single example. }
    \Description{Frequency of each method from the data generators}
    \label{fig:frequencies}
\end{figure}

Figure~\ref{fig:frequencies} shows the distribution of optimal methods based on this metric, i.e. how often a method produces the smallest DAG size. The dataset is imbalanced: this is attributed to the top methods being used more generally whilst the less frequently used methods are for more niche cases.  For details on each individual method see the Maple help page: \url{https://www.maplesoft.com/support/help/maple/view.aspx?path=int%2Fmethods}

\subsection{Data Preprocessing}
\label{sec:data/preprocessing}

We next tokenize the data:  we create tokens corresponding directly to the functions, operators, and variables which appear.  For numbers, we start by including a flag token, similar to Lample \& Charton \cite{Lample2020}:  \texttt{INT+} or \texttt{INT-} to denote whether the coming integer being encoded is positive or negative. However, we then differ from \cite{Lample2020} and avoid creating separate digits for each token.  Instead, we encode the remaining (positive) integer as follows:  
\begin{enumerate}
    \item if the number is $0, 1 \text{ or } 2$, then we create a corresponding token;
    \item if the number is a single-digit other than $0, 1 \text{ or } 2$, we replace the number with a \verb|CONST| token;
    \item if the number has two digits, we replace by a \verb|CONST2| token; and
    \item for all other cases, we replace by a \verb|CONST3| token.
\end{enumerate}
Unlike \cite{Lample2020}, we are not generating output integrands directly but instead selecting methods: the specific values of constants are generally insignificant in deciding which method to use.  This approach allows us to simplify the data without losing essential information and should aid in the generalisation on data out of distribution. We retain integers within the range $[-2,2]$ as they can influence integration outcomes (e.g. $(x+1)^{1}$ and $(x+1)^{-1}$ integrate very differently). 

To maintain dataset diversity, we eliminate duplicate integrands after the transformations with the \verb|CONST| tokens that arise, so the dataset comprises unique expressions only. Additionally, following \cite{Devlin2019_bert}, we prepend each sequence with a \texttt{[CLS]} token, which serves as a single representation of the entire sequence/tree. The final encoded output of this \texttt{[CLS]} token is then used for the actual classification.

\subsection{Training and Testing Datasets}
\label{subsec:traintest}

We produce data from the six different data generation methods described in Section~\ref{sec:data/generation}:  FWD, BWD, IBP, RISCH, SUB, and LIOUVILLE, and then label and pre-process the data as described above.  

We construct a training dataset of 1,000,000 examples.  The labelling process was extremely time-consuming, but it is a cost that only has to be paid once.  We generate using an even split between all six data generators. 

We also construct a separate hold-out test set from the data generators . The holdout test set is split between 50,000 elementary expressions coming from the six data generators and 20,000 non-elementary expressions coming only from the FWD, BWD, and LIOUVILLE generators.

\subsection{Independent Validation Dataset}

We also utilise an entirely separate dataset, the ``\emph{Maple Test Suite}''.  This is maintained by Maplesoft and comprises of unit tests, user queries and bug reports collected over the years to ensure that Maple's integration procedure works properly. The Maple test suite is critical to the evaluation of our work:  we do not use it for training, keeping it separate to it may be considered as entirely independent OOD data. Performance on this dataset empirically demonstrates an ML model's ability to generalise. 

To match the format of the training data, we make a few modifications to the raw data.
\begin{itemize}
    \item We filter the dataset to only consider expressions that have functions that are part of the training vocabulary.
    \item The variable of integration may be different than $x$. If this is the case, we perform a substitution so that the variable of integration is $x$ as in the training data.
    \item Some expressions have symbolic coefficients (e.g. $ax^2 + b$). As none of our training data had any symbolic coefficients, we opt to replace them with a \verb|CONST3| token.  We choose this over \verb|CONST1| or \verb|CONST2| guided by Maplesoft who report that in general parametric methods will give an answer with the same shape as that for a large integer value of the parameter.  
    \item Sometimes, this replacement may create a division by $0$ error (e.g. $\frac{1}{(a-b)x}$), so we then vary the substitution between \verb|CONST2| and \verb|CONST3| if this happens.
\end{itemize}
After filtering the dataset, we have 13,040 examples to validate on (of which 2,980 contain non-elementary functions)

\section{Model architectures for Sequences and Trees}
\label{sec:architecture}

We experiment with LSTMs and transformers: deep learning models used in many natural language processing tasks such as language translation or sentiment analysis. Here, we compare the difference between the versions of these models that receive their input as a sequence, and those that receive their input as a tree.   

\subsection{Sequence-Based Models}
\label{sec:architecture/seq}

As we saw in Figure~\ref{fig:representations}, one of the possible ways to represent a mathematical expression is sequentially in prefix notation. Each operand and operator is represented as its own token, and is naturally fed as input data to ML models that accept sequences of tokens. In general, a sequence of tokens could come from text, speech, or time series data. 

\begin{figure}[ht]
    \centering
    \includegraphics[width=0.4\linewidth]{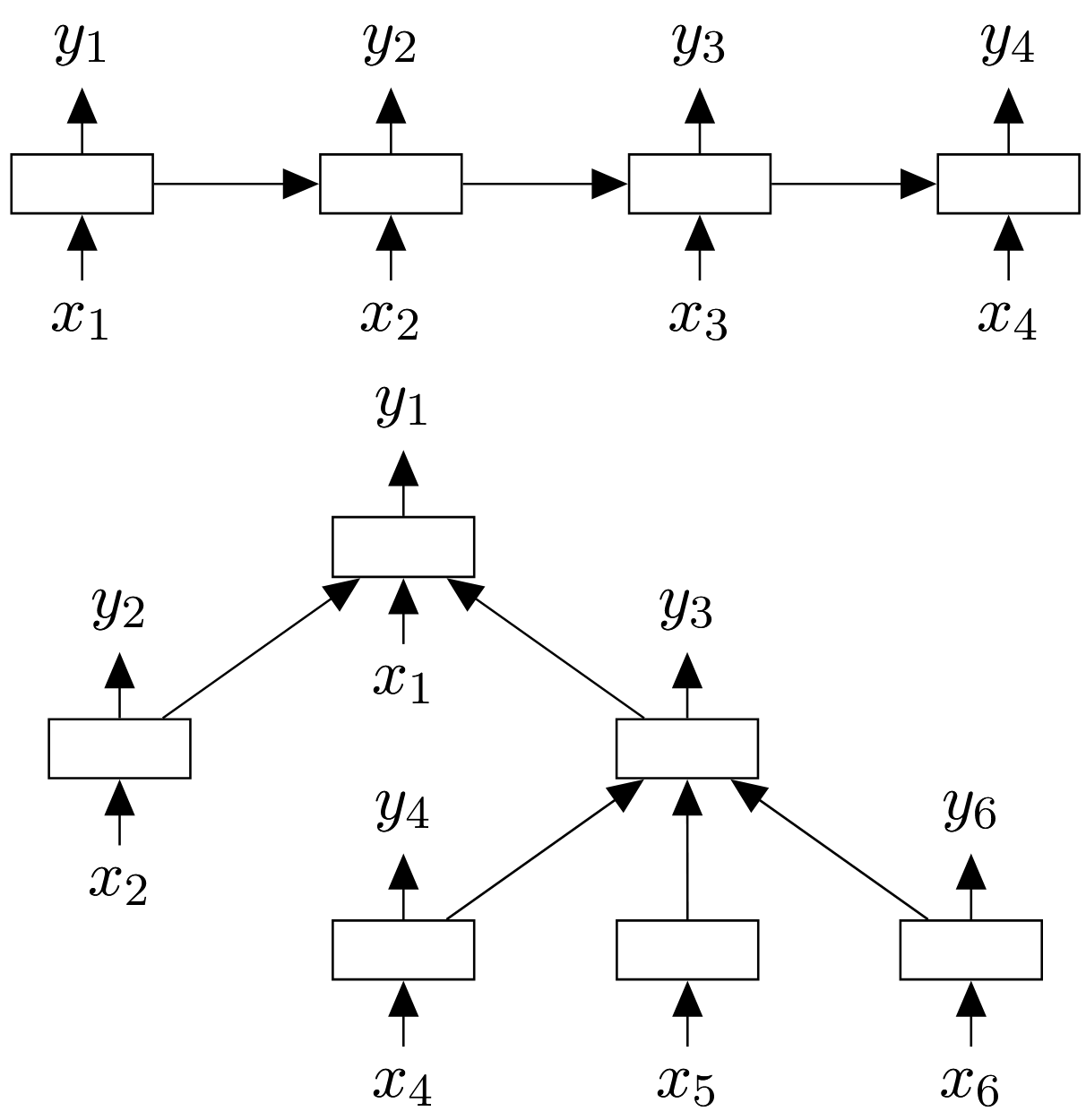}
    \caption{A comparison of a sequence-based LSTM (top) and a TreeLSTM (bottom) as described in \cite{Tai2015_TreeLSTM}.}
    \Description{LSTM and TreeLSTM}
    \label{fig:TreeLSTM}
\end{figure}

One ML model to process sequence data is the Long Short-Term Memory (LSTM) \cite{Horchreiter1997_LSTM}, a type of recurrent neural network that is particularly good at handling the vanishing gradient problem (when gradients become very small during back propagation). LSTMs are designed to remember information over long sequences, based on a cell that has three special ``gates'' to achieve this.
\begin{itemize}
    \item \textbf{Forget gate}: decides what old information to discard.
    \item \textbf{Input gate}: decides what new information from the current token to add.
    \item \textbf{Output gate}: decides which information to pass to the next cell.
\end{itemize}
Each cell takes an input token $x_i$ and process it along with the information from tokens $x_0, \ldots, x_{i-1}$ (via the state of the previous cell); and then pass this information along to the cell processing token $x_{i+1}$. This is visualised in the top part of Figure~\ref{fig:TreeLSTM}. The issue with LSTMs is that relationships between tokens near the beginning and the end of the sequence are harder for the model to grasp. Furthermore, LSTMs cannot process information in parallel since the current cell is dependent on the state of the previous cell, meaning training on a large amount of data is time-consuming.

\begin{figure}[h]
    \centering
    \includegraphics[width=0.2\linewidth]{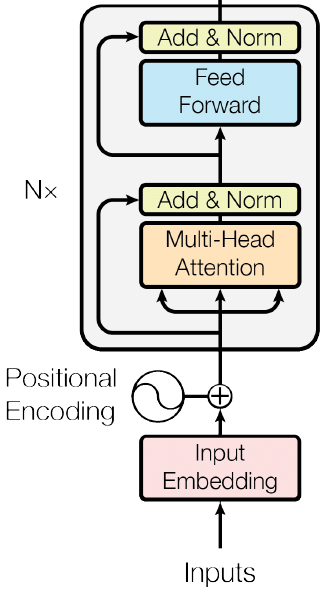}
    \caption{The encoder architecture of a transformer \cite{Vaswani2017_transformer}. The importance of the positional encoding gives order to the sequence of tokens. There is normally also a decoder architecture, but only the encoder is needed to make predictions in this task.}
    \label{fig:transformer}
    \Description{The architecture of an encoder in a transformer. Input embeddings and positional embeddings are added together, and then we calculate attention across multiple heads. The final output of the attention mechanism is fed into a feed-forward network to produce the final embeddings.}
\end{figure}

Transformers, introduced in \cite{Vaswani2017_transformer}, fix both of these issues by being able to ``look'' at the entire sentence at once. This is done by the attention mechanism, which allows each token to focus on the important/relevant tokens in the rest of the sequence to gain understanding. This gives it the ability to understand the relationship of tokens further apart in the sequence, which LSTMs struggle with. In addition, attention is calculated between pairs of tokens, meaning that this can be done in parallel, allowing for faster training and thus training with more data than LSTMs. The attention mechanism essentially treats the input tokens as a bag of words rather than a sequence, and acting along it has no knowledge of the order tokens are presented in. Transformers thus also include a positional encoding step to include information about the order of the sequence: this takes the form of a vector added to each token embedding that contains information about its position, as illustrated in Figure \ref{fig:transformer}. Transformers were the architecture used by Lample and Charton \cite{Lample2020}, and have since been used in many symbolic mathematics tasks, such as symbolic regression (predicting the mathematical expression of a function from observations of its values) \cite{Kamienny2022, d2ascoli2023ODE}.

\subsection{Tree-Based Models}
\label{sec:architecture/tree}

Many researchers in the area of AI for mathematics have noted that mathematical expressions can be represented naturally as unary-binary trees, but very few have actually leveraged this information in ML model architecture. We know only of \cite{Wang2021_mathembed}, as discussed in Section~\ref{subsec:LRTree}.  
We will compare the standard sequential LSTMs and transformers described above to tree-based variants:  TreeLSTMs \cite{Tai2015_TreeLSTM} and Tree Transformers \cite{Shiv2019_TreeTransformer}. 

TreeLSTMs generalise an LSTM cell to a tree structures by computing each node’s hidden state and cell state as functions of its children’s states, rather than just linear sequences \cite{Tai2015_TreeLSTM}. This is visualised in the bottom of Figure~\ref{fig:TreeLSTM}. In the context of mathematical expressions, the operands would pass their information to the parent operator to produce the hidden state of a sub-expression. For example, in the expression $\sin(x^2)$, $x$ and $2$ would aggregate their hidden states to the power operator $\wedge$, and the sub-expression $x^2$ would pass its hidden state to the function operator $\sin$.   

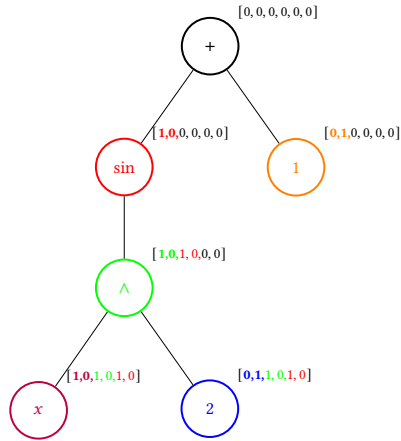
\begin{figure}[ht]
  \centering
  \resizebox{0.35\textwidth}{!}{%
    \begin{forest}
      for tree={
        circle, draw,
        line width=1pt,
        minimum size=1cm,
        inner sep=1pt,
        s sep=2cm,    
        l sep=1.1cm,    
        label distance=1.5ex
      }
      [$+$, label=above right:{\footnotesize$[0,0,0,0,0,0]$}
          [\textcolor{red}{$\sin$}, 
             draw=red, label=above right:{\footnotesize$[\textcolor{red}{\textbf{1,0,}} 0,0,0,0]$}
            [\textcolor{green}{$\wedge$}, 
               draw=green, label=above right:{\footnotesize$[\textcolor{green}{\textbf{1,0,}}\textcolor{red}{1,0,} 0,0]$}
              [\textcolor{purple}{$x$}, 
                draw=purple, label=above right:{\footnotesize$[\textcolor{purple}{\textbf{1,0,}}\textcolor{green}{1,0,}\textcolor{red}{1,0}]$}
              ]
              [\textcolor{blue}{$2$}, 
                draw=blue, label=above right:{\footnotesize$[\textcolor{blue}{\textbf{0,1,}}\textcolor{green}{1,0,}\textcolor{red}{1,0}]$}
              ]
            ]
          ]
          [
          \textcolor{orange}{$1$}, 
          draw=orange, label=above right:{\footnotesize$[\textcolor{orange}{\textbf{0,1,}} 0,0,0,0]$}
          ]
    ]   
    \end{forest}%
  }
  \caption{Syntax tree for $\sin(x^2)+1$ with each node’s tree transformer positional encoding to its right. For each new depth, you push an additional size 2 one-hot vector into the positional encoding for the node, indicating whether it is the left child or right child.}
  \label{fig:sin-tree-positional}
  \Description{Example of tree encoding}
\end{figure}

The Tree Transformer modifies the positional encoding mechanism discussed above to explicitly represent where each node is within the tree \cite{Shiv2019_TreeTransformer}. In regular (sinusoidal) positional encoding, each token in the sequence has a unique position vector encoded with sine and cosine functions, providing a smooth continuous embedding. Like regular positional encodings, tree transformers encode a unique position vector for each token, but this now encoded parent and child information using a one-hot encoding which read from left to right follows the tree up from the current node.  Note that the encoding has a fixed depth which may be shallower than the actual tree.  An example is shown in Figure~\ref{fig:sin-tree-positional}.

\subsection{Maple's Baseline Integrator}

\begin{figure}[ht]
    \centering
    \includegraphics[width=0.8\linewidth]{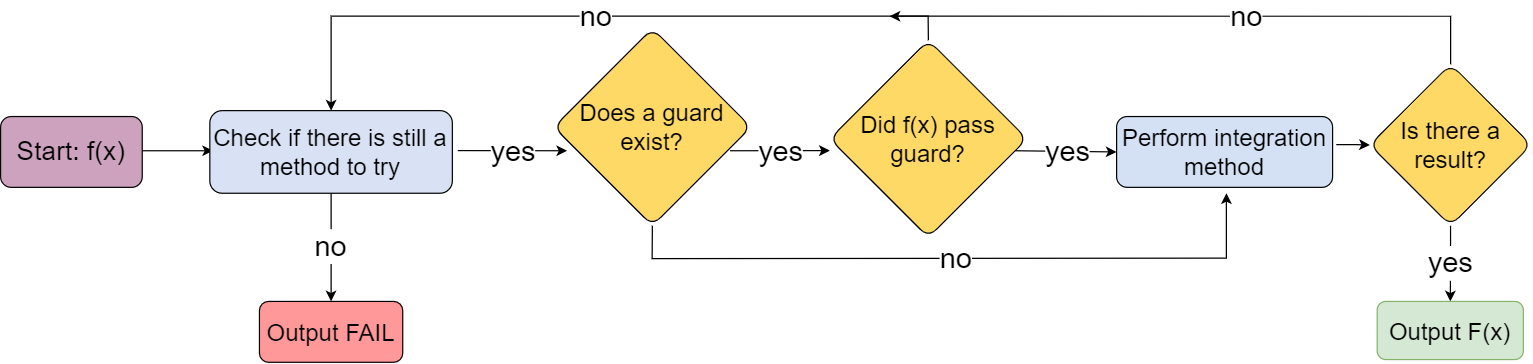}
    \caption{A high-level overview of how the indefinite integral command works in Maple to calculate $F(x) = \int f(x) \text{dx}$.}
    \label{fig:maple_int}
    \Description{A flowchart of Maple's integration method selector. We first check if we can do a method, and it passes its guard. If we perform the method and it fails, we try again with another method until either an answer is produced or all methods are exhausted.}
\end{figure}

We will compare the tree-based variants of the aforementioned models to their sequential counterparts. In addition, we will also compare against Maple's existing integrator as a baseline.  This will try each method in a fixed order until a method successfully outputs an integral. Maple's integrator has been hand-tuned by mathematicians over several decades, so this order is well chosen but the order does not change for a particular problem instance as the ML approaches will.  Maple's baseline method does include, for some of the methods, what is known as a ``\emph{guard}'': simple heuristic code (often a type check) to decide whether a method can be applied or not (if not, that method is skipped in the sequence). The guards in Maple are our inspiration for the classification stage in our two-stage framework that we will shortly discuss in Section \ref{sec:methodology/ranking}. A flowchart describing Maple's method selector is shown in Figure \ref{fig:maple_int}.  

\section{Problem Framing}
\label{sec:methodology}

In addition to the model architecture, how the problem is framed for ML plays a crucial role in the quality of the results. We discuss the different framings we have considered in this section.  

\subsection{Binary Classification}
\label{sec:methodology/classification}

Perhaps the simplest framing is to view this as a classification problem, where we seek to classify a problem instance by which method is best suited for it. We explored this in our preliminary work \cite{Barket2024_TreeLSTM}, and give a high-level overview here. As there are multiple methods that can produce the optimal answer, this is framed as a multi-label classification problem. The most intuitive way to handle multi-label classification is binary relevance \cite{Zhang2018_BR}: if $L$ is the set of methods, then $|L|$ models are trained for predicting binary classification on each method. Essentially, each binary classifier is predicting whether or not the method used will produce the smallest DAG size or not.  

In \cite{Barket2024_TreeLSTM}, this framework worked well on test data that came from the same data generators as the training data, but the model(s) did not perform well on OOD data. There are multiple possible reasons for weak generalisation. First, the data we are training on is imbalanced between the methods. As shown in Figure~\ref{fig:frequencies}, some methods have many more positive samples than others and so models may just be learning this bias. Secondly, models are only trained on the best method to use, and methods that are close to the best are ignored when in reality we have a strong preference between using a close second and a very poor choice. Lastly, the models cannot learn any correlations between the methods during training as each classifier is trained independently. We could include such information into this framework by using classifier chains \cite{Read2009_CC}, where the models are trained in turn with each subsequent model including the output of the previous models as part of the input. However, that approach has its own drawbacks, such as introducing sensitivity in the order of model training and error propagation passed on to each model if earlier models are trained incorrectly. Instead, we aim to address these issues by switching the problem framework. 

Our first attempt was to address this by switching from classification to regression: having the models predict the DAG size so that we can select the one with the lowest prediction, as was used in \cite{Del2024_lessons} for another computer algebra optimisation task.  We report in Appendix~\ref{sec:regression} how this led to only modest improvements and had its own drawbacks that led us to try the next problem formulation instead.    

\subsection{Two-Stage Classification + Ranking Framework}
\label{sec:methodology/ranking}

\begin{figure}[h]
    \centering
    \includegraphics[width=0.95\linewidth]{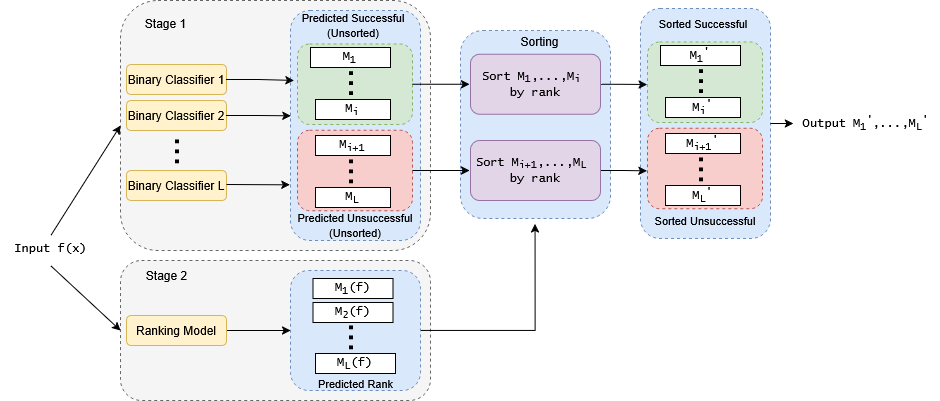}
    \caption{The two-stage classification + ranking architecture. Here, $\text{M}_i$ represents integration method $i$ (of $|L|$ possible methods) and $\text{M}_i(f)$ is the rank given to $\text{M}_i$ given input expression $f(x)$. The output is the order of methods to try.}
    \label{fig:architecture}
    \Description{two-stage architecture}
\end{figure}

Considering the drawbacks of the previous frameworks, we switch the paradigm to a two-stage \emph{classification + ranking} framework, as visualised in Figure~\ref{fig:architecture}. The first stage is predicting whether the model will \textit{succeed} at integration or not,  while the second stage \textit{ranks} the methods from best to worst, ensuring all models predicted to succeed are ranked above those predicted to fail.

The classification stage is again multi-label classification using the use the binary relevance method with $|L|$ binary classifiers as above, but this time we are not classifying whether a method leads to the minimum DAG size, just whether it can produce an answer at all.  These thus play the same role as the guards in the current Maple method selector.  This stage is mainly to help with reducing computation time by making sure that we do not try methods that are likely going to end up failing.  We explored the training of such models in a preliminary work \cite{Barket2024_guards} and we discuss their role further later in Section~\ref{sec:results/2stage_analysis}. 

The second stage ranks the methods using a masked and weighted version of the RankNet loss \cite{Burges2005_RankNet} metric:
\begin{align*}
\mathcal{L}_{\mathrm{rank}} &=
  \sum_{j=i+1}^{|L|} \sum_{i=1}^{|L|}
  \delta_{ij}\, w_{ij}\, \ln\left(1 + e^{p_i - p_j}\right) && \\
\text{where } \delta_{ij} &=
\begin{cases}
1 & \text{if } m_i = m_j = 1 \text{ and } y_i < y_j \\
0 & \text{otherwise}
\end{cases} && \\
\text{and } w_{ij} &= \frac{1}{\log_2\!\left(\tfrac{y_i + y_j}{2} + 2\right)} &&
\end{align*}
Here, $L$ is the set of methods, and we iterate over each possible pair of methods. For each label $i$, the ground-truth rank is $y_i$, the model's predicted score is $p_i$, and the mask $m_i$ is 1 if label $i$ is present, and 0 otherwise. In simpler terms, we are doing pairwise comparisons between pairs of methods, and the model incurs more loss when the order between the two compared methods is incorrect (i.e. the model predicted a method with a larger DAG size did better than a method with a smaller DAG size). The mask is to ensure that we only compare methods that produced an answer in the training data.   

The rank-based weight $w_{ij}$ is adapted from normalised discounted cumulative gain measure discounting \cite{Jarvelin2002_NDGC}. The weighting places more importance on the correct ordering of highly-ranked labels (i.e. has the smallest DAG size). We define the pair mask $\delta_{ij}$ to select only those pairs where both labels are present and the $y_i$ ranks higher than $y_j$.  

Switching from purely classification to a two-stage classification + ranking framework helps with many of the problems stated before. There is less of an issue with data imbalance since we are training one singular model rather than $|L|$ different classifiers. In classification, we would punish all methods that were not the best. Now, we can more accurately predict which methods came in $2^{\text{nd}}$, $3^{\text{rd}}$, etc. This is beneficial in that if we end up misclassifying the best method, the model will have a better chance at predicting a method that is close to the best answer, and has the least likelihood of predicting the worst method. Lastly, the benefit of having a single ranker is that the model can learn the correlations between methods directly; there is no need for techniques such as classifier chains to learn label correlations.

\section{Results}
\label{sec:results}

We now describe our empirical evaluation of all the models and frameworks introduced in Sections~\ref{sec:architecture} and \ref{sec:methodology} at the task of selecting the best method based on DAG size. The four model architectures will be run in each of the two frameworks, and will also be compared to Maple's current method selector, acting as a baseline.

All ML models are trained using the same training dataset described in Section~\ref{subsec:traintest}. The models and the Maple selector are evaluated on both the hold-out test dataset (Section~\ref{subsec:traintest}) of 70,000 examples, and the independent Maple test suite (Section~\ref{sec:data/preprocessing}) of 13,040 examples.

\subsection{Binary Classification}
\label{sec:results/classification}

\begin{figure}[htbp]
  \centering
  \begin{subfigure}[b]{0.65\textwidth}
    \includegraphics[width=\textwidth]{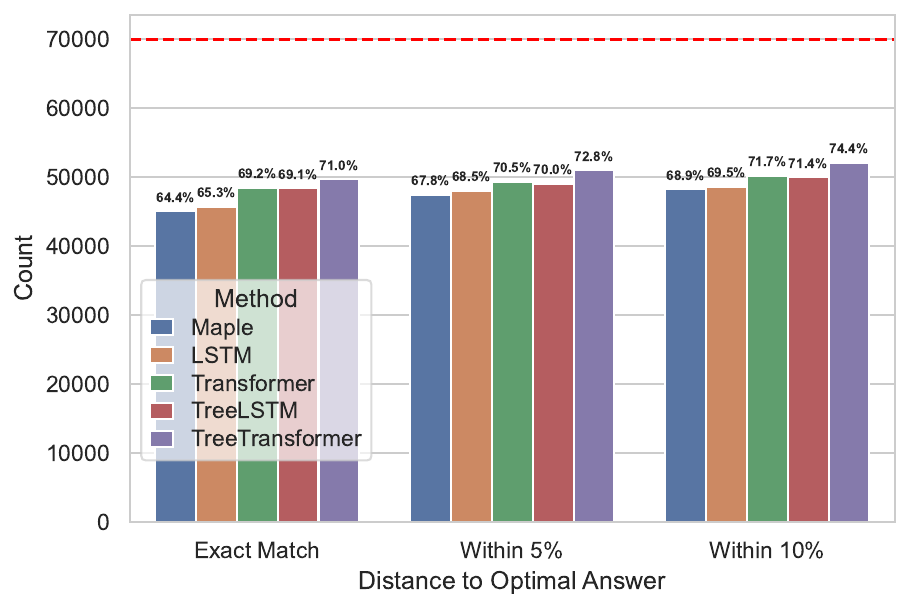}
    \caption{Binary classification results on the hold-out test set}
    \label{fig:results/bin_cls/test}
  \end{subfigure}
  \hfill
  \begin{subfigure}[b]{0.65\textwidth}
    \includegraphics[width=\textwidth]{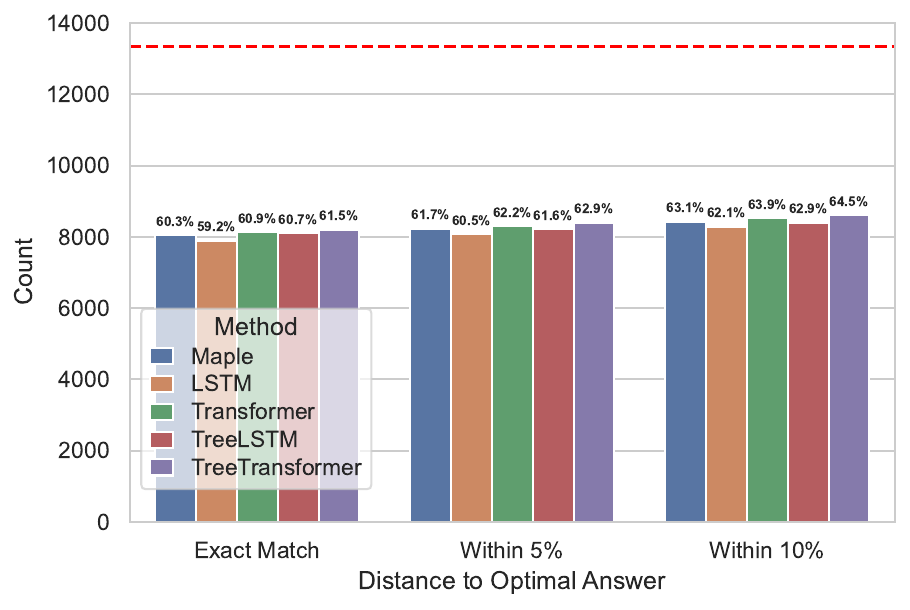}
    \caption{Binary classification results on the Maple Test Suite}
    \label{fig:results/bin_cls/maple}
  \end{subfigure}
  \caption{Results of multi-label binary classification.  \label{fig:results/bin_cls}}
  \Description{Binary Classification Results}
\end{figure}

We first evaluate performance under the binary classification framework discussed in Section \ref{sec:methodology/classification}.  We reported on the performance of LSTMs and TreeLSTMs under this framework in the preliminary work  \cite{Barket2024_TreeLSTM}. The numbers are different here as the dataset has been updated to include data from new generators and data with non-elementary functions.  We also report for the first time on the performance of transformer models.  The results are summarised in Figure~\ref{fig:results/bin_cls}(a) for the hold-out test suite and Figure~\ref{fig:results/bin_cls}(b) for the Maple test-suite.  In each case the red dotted line indicates the total dataset size, and the bars on the left how many of those problems each model was able to select the optimal method for.  We also report the number of problems for which models got within 5\% and 10\% of the optimal answer.

We see the models do well when evaluated on test data generated similarly to the training data, but that performance drops off significantly when we evaluate on the independent Maple test suite.  Here, only the tree versions of the models outperform the baseline Maple selector, and even then only by a small number of problems.  This indicates the models have weak generalisation abilities. Another important note is the weak gains when looking at the close-to-optimal answers. There is only a very small increase in performance when evaluating with this more forgiving metric.

\subsection{Classification + Ranking}
\label{sec:results/2stage}

\begin{figure}[htbp]
  \centering
  \begin{subfigure}[b]{0.65\textwidth}
    \includegraphics[width=\textwidth]{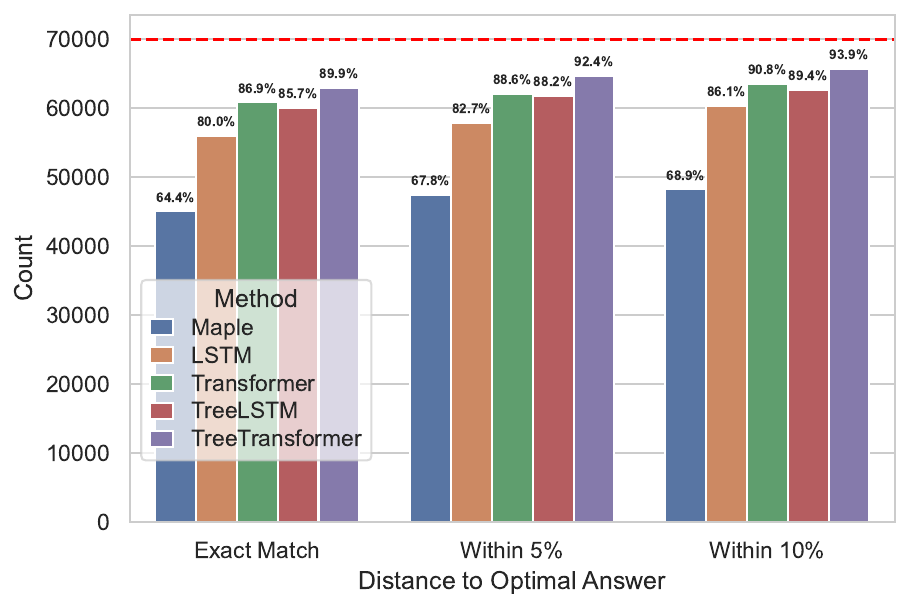}
    \caption{Two-stage results on the hold-out test set}
  \end{subfigure}
  \hfill
  \begin{subfigure}[b]{0.65\textwidth}
    \includegraphics[width=\textwidth]{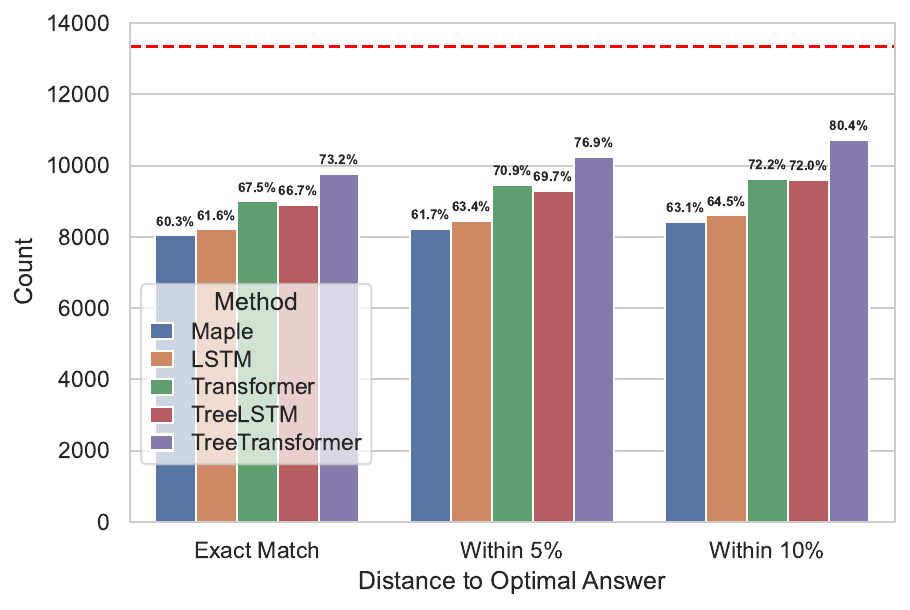}
    \caption{Two-stage results on the Maple Test Suite}
  \end{subfigure}
  \caption{Results of the two-stage classification + ranking framework. \label{fig:results/2stage} }
  \Description{classification and ranking results}
\end{figure}

We now evaluate the two-stage framework (Section~\ref{sec:methodology/ranking}): the results are in Figure~\ref{fig:results/2stage}, in the same format as Figure~\ref{fig:results/bin_cls}.

We see a significant improvement in the results on the hold-out test set, with the tree transformer now predicting the best method almost 90\% of the time. On the independent Maple test suite, the models are still not generalising perfectly, with the tree transformer performance falling to 73\% on exact matches. However, this is a big improvement compared to binary classification in Figure~\ref{fig:results/bin_cls/maple}, where the model could barely improve on the baseline Maple's selector. We may conclude that we have improved upon the state-of-the-art, and crucially that we do so on independent OOD data. 

Important to note is that we see a much better improvement on the near match evaluation compared to when we used just classification. The tree transformer performance rises to 80\% when we allow output within 10\% of the optimum, 17.3 points above the baseline. This shows the importance of ranking: if a model fails to select the best method all is not lost as it has been trained to rank the methods and so can pick the next best accordingly. 

Another interesting point is the comparison of exact matches on the elementary versus non-elementary cases. Although integrating a non-elementary expression is considered harder mathematically, diving into the results we find that the ML accuracy on exact matches for non-elementary expressions is actually greater than the elementary ones (e.g. $92.7\%$ versus $88.1\%$ for the tree transformer). We think this is because there are fewer applicable methods to use for non-elementary expressions compared to elementary ones: on average, an elementary example has 2.9 possible applicable methods whereas a non-elementary example has 1.7 possible applicable methods. This thus makes the learning task easier, despite there being less data in the training set.    

\subsection{Tree versus Sequential Architecture}

Note that in all comparisons of the models across Figures~\ref{fig:results/2stage} and \ref{fig:results/bin_cls} the tree-based variant always does better than its sequential counterpart. There is a modest improvement between the tree and regular transformers, and a more substantial gap between the tree and regular LSTMs. This shows a clear benefit of representing a mathematical expression explicitly as a tree for this task.

\subsection{Some examples of different answers from different methods}
\label{sec:results/compare}

Table~\ref{tab:answer_compare} shows the difference in answers between a tree transformer guided answer and what Maple currently outputs for some hand-picked examples from the Maple test suite. We observe that the difference goes far beyond simple simplification from bracket expansion or the application of trigonometric identities. Consider the example $\frac{\ln(x^4+1)}{x}$: both possible answers introduce new functions (polylog and dilog respectively), but the current Maple answer also introduces irrational and complex numbers, as well as being much longer overall. We note that there are other examples we could have picked with even more dramatic differences between the Maple answer and the tree transformer answer, but they would take multiple pages to include! 

The final example in the table was included to demonstrate that our DAG size metric does not always align with expression length. Although the ML answer looks slightly longer, its DAG representation in Maple is smaller than Maple's answer (since the $x^i$ sub-expression is stored only once). 

The reader may wonder if we can just use computer algebra simplification tools to take the non-optimal expressions and find the optimal ones.  In Table~\ref{tab:answer_compare}, this would only be achieved for the first example. Computer algebra simplification is far from straightforward \cite{Carette2004, Stoutemyer2011, CDJLW02}, and while it can be useful it is preferable for an algorithm to produce the simplest output directly if possible.  

\begin{table}[htp]
    \centering
    \renewcommand{\arraystretch}{3.27} 
    \caption{The difference between Maple's suggested answer and our Tree Transformer suggested answer for some hand-picked examples. Note that the answers in each row are mathematically equivalent.}
    \label{tab:answer_compare}
    \begin{tabular}{|c|>{\centering\arraybackslash}p{7.15cm}|>{\centering\arraybackslash}p{2.3cm}|}
\hline
\textbf{Integrand} & \textbf{Maple Result} & \textbf{Tree Transformer}\\
\hline

$r^2\cos(x)^2 + r^2\sin(x)^2$ & $\displaystyle r^2\left( \frac{\cos(x)\sin(x)}{2} + \frac{x}{2} \right) + r^2\left( - \frac{\cos(x)\sin(x)}{2} + \frac{x}{2} \right)$ & $r^2x$\\

\hline

$\displaystyle \frac{\ln(x^4 + 1)}{x}$ & $\ln \! \left(x \right) \ln \! \left(x^{4}+1\right) - \ln \! \left(x \right) \ln \! \left(\frac{\frac{\sqrt{2}}{2}+\frac{\mathrm{i} \sqrt{2}}{2}-x}{\frac{\sqrt{2}}{2}+\frac{\mathrm{i} \sqrt{2}}{2}}\right)-\text{dilog}\! \left(\frac{\frac{\sqrt{2}}{2}+\frac{\mathrm{i} \sqrt{2}}{2}-x}{\frac{\sqrt{2}}{2}+\frac{\mathrm{i} \sqrt{2}}{2}}\right)-\ln \! \left(x \right) \ln \! \left(\frac{-\frac{\sqrt{2}}{2}+\frac{\mathrm{i} \sqrt{2}}{2}-x}{-\frac{\sqrt{2}}{2}+\frac{\mathrm{i} \sqrt{2}}{2}}\right)-\text{dilog}\! \left(\frac{-\frac{\sqrt{2}}{2}+\frac{\mathrm{i} \sqrt{2}}{2}-x}{-\frac{\sqrt{2}}{2}+\frac{\mathrm{i} \sqrt{2}}{2}}\right)-\ln \! \left(x \right) \ln \! \left(\frac{-\frac{\sqrt{2}}{2}-\frac{\mathrm{i} \sqrt{2}}{2}-x}{-\frac{\sqrt{2}}{2}-\frac{\mathrm{i} \sqrt{2}}{2}}\right)-\text{dilog}\! \left(\frac{-\frac{\sqrt{2}}{2}-\frac{\mathrm{i} \sqrt{2}}{2}-x}{-\frac{\sqrt{2}}{2}-\frac{\mathrm{i} \sqrt{2}}{2}}\right)-\ln \! \left(x \right) \ln \! \left(\frac{\frac{\sqrt{2}}{2}-\frac{\mathrm{i} \sqrt{2}}{2}-x}{\frac{\sqrt{2}}{2}-\frac{\mathrm{i} \sqrt{2}}{2}}\right)-\text{dilog}\! \left(\frac{\frac{\sqrt{2}}{2}-\frac{\mathrm{i} \sqrt{2}}{2}-x}{\frac{\sqrt{2}}{2}-\frac{\mathrm{i} \sqrt{2}}{2}}\right)
$ & $\displaystyle \frac{\text{polylog}(2,-x^4)}{4}$\\

\hline

$-{\mathrm e}^{2 \,\mathrm{i} \pi  x} \left(\sin \! \left(2 \pi  x \right) \mathrm{i}-\cos \! \left(2 \pi  x \right)\right)$ & $\displaystyle \frac{xe^{2\mathrm{i}\pi x} - xe^{2\mathrm{i}\pi x}\tan(\pi x)^2 - 2ixe^{2\mathrm{i}\pi x} \tan(\pi x) }{1 + \tan(\pi x)^2}$ & $x$\\

\hline

$\text{AiryAi}(x)^2$ & $\displaystyle -\frac{18^{2/3}2^{1/3}3^{2/3}x^4\text{BesselI}\left(\frac{2}{3},\frac{2x^{3/2}}{3}\right)^2}{162(x^{3/2})^{4/3}} + \frac{2\cdot18^{2/3}2^{1/3}3^{2/3}x\text{BesselI}\left(\frac{2}{3},\frac{2x^{3/2}}{3}\right)^2}{81(x^{3/2})^{4/3}} + \frac{18^{2/3}2^{1/3}3^{2/3}x^4\text{BesselI}\left(\frac{5}{3},\frac{2x^{3/2}}{3}\right)^2}{162(x^{3/2})^{4/3}} - \frac{18^{2/3}2^{1/3}3^{2/3}\sqrt{x}\text{BesselI}\left(\frac{1}{3},\frac{2x^{3/2}}{3}\right)\text{BesselI}\left(\frac{2}{3},\frac{2x^{3/2}}{3}\right)}{81} - \frac{18^{2/3}2^{1/3}3^{2/3}\text{BesselI}\left(\frac{1}{3},\frac{2x^{3/2}}{3}\right)\text{BesselI}\left(\frac{5}{3},\frac{2x^{3/2}}{3}\right)}{81} + \frac{18^{2/3}2^{1/3}3^{2/3}x^2\text{BesselI}\left(\frac{4}{3},\frac{2x^{3/2}}{3}\right)\text{BesselI}\left(\frac{2}{3},\frac{2x^{3/2}}{3}\right)}{81} + \frac{3^{5/6}2^{2/3}\Gamma\left(\frac{5}{6}\right)x^3\text{hypergeom}\left([\frac{5}{6},1],[\frac{4}{3},\frac{5}{3},2],\frac{4x^3}{9}\right)}{36\pi^{3/2}}$ & $\text{AiryAi}(x)^2x - \text{AiryAi}(1,x)^2$\\

\hline

$\displaystyle \frac{\sin(\ln(x))}{x}$ & $-\cos(\ln(x))$ & $\displaystyle -\frac{x^{\mathrm{i}}}{2} - \frac{1}{2x^{\mathrm{i}}}$ \\
\hline
\end{tabular}
\end{table}

\subsection{Misclassification Analysis}

\begin{figure}[ht]
    \centering
    \includegraphics[width=0.6\linewidth]{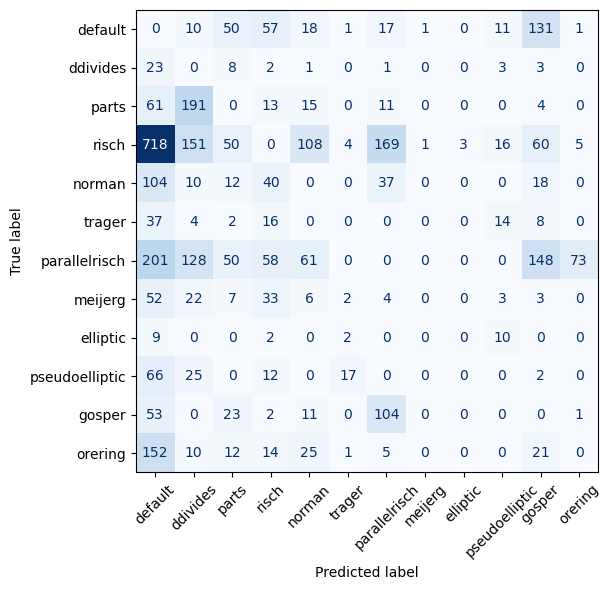}
    \caption{Confusion matrix of examples that did not have the correct method predicted from the Maple test suite.}
    \label{fig:confusion_matrix}
    \Description{The confusion matrix of examples from the Maple Test Suite. This means that for the methods that are listed as the true prediction, the number says how many went to the predicted method.}
\end{figure}

To understand how the ML models may be improved, we take a closer look at some examples from the Maple test suite where the Ml was incorrect: Figure \ref{fig:confusion_matrix} shows the confusion matrix for those examples that did not have the best method correctly predicted by the Tree Transformer. One glaring issue is the model's bias in predicting the \texttt{default} method. When the model makes an incorrect prediction, it is most likely predicting \texttt{default} rather than the correct method. This should not come as a huge surprise: Figure \ref{fig:frequencies} showed that \texttt{default} is one of the most commonly successful methods, and unlike all the others, \texttt{default} is designed to be a general purpose integrator, applicable to all problems. This means it is harder to learn the patterns in the dataset that have \texttt{default} as the optimal method. This could be addressed by penalising the model more when it incorrectly predicts \texttt{default}, but there is no guarantee the model would then select the optimal methods instead.

The methods whose true labels are most commonly misclassified in the Maple test suite are \texttt{risch} and \texttt{parallelrisch}. This is also unsurprising as these two methods are the two most commonly optimal in the dataset (see again Figure \ref{fig:frequencies}). Examples for the \texttt{risch} method get misclassified for \texttt{default} the majority of the time, but the \texttt{parallelrisch} mispredictions have a roughly equal spread over many of the methods. Again, we could address the former by re-weighting \texttt{default} so that choosing this incorrectly is penalised more. Fixing the latter would be harder, as we would need to understand why each example is getting misclassified for all the different methods. 

Some other interesting points to draw from Figure \ref{fig:confusion_matrix}:
\begin{itemize}
    \item \texttt{gosper} is getting predicted often despite being one of the least common methods for optimality.
    \item The majority of the time that \texttt{gosper} is misclassified, it is for \texttt{parallelrisch}.
    \item \texttt{risch} also gets mistaken for \texttt{parallelrisch} often, but this is not the case in the other direction.
    \item The majority of the time that \texttt{parts} is misclassified, it is for  \texttt{derivativedivides}. 
\end{itemize}

A major difference between the testing dataset and the Maple test suite is the inclusion of symbolic coefficients in the latter. As mentioned in Section \ref{sec:data/preprocessing}, our training data did not have any expressions with symbolic coefficients and this issue was resolved by replacing the coefficients with \verb|CONST| tokens. However, it makes a lot more sense to generate expressions with symbolic coefficients in the training data. Of the examples that are misclassified, 57\% had such symbolic coefficients, which further reinforces the fact that generating examples with symbolic coefficients should be beneficial.

We show some hand-picked examples from the Maple Test Suite where the Tree Transformer failed to pick the best answer in Table \ref{tab:answer_compare_bad}. Note that although the Tree Transformer picked a sub-par method for these, it never picked the worst one (the method with the largest DAG size). Of the ones that are misclassified in the Maple test suite, only 26.5\% of those examples had the worst possible method chosen. This further highlights the importance of ranking rather than just classifying. 
Furthermore, of those 26.5\% examples where the worst method was selected, 67.5\% only had two possible methods to select from, meaning that the model could only select between the best method and the worst method with no middle-ground options, i.e. the ranking methodology could offer no further help for these.

\begin{table}[ht]
    \centering
    \renewcommand{\arraystretch}{3}
    \caption{Examples where the Tree Transformer did not select the best method for the given integrand.}
    \label{tab:answer_compare_bad}
    \begin{tabular}{|c|>{\centering\arraybackslash}p{7cm}|>{\centering\arraybackslash}p{4.1cm}|}
\hline
\textbf{Integrand} & \textbf{Tree Transformer} & \textbf{Best Answer}\\
\hline
$\displaystyle \frac{\mathrm{arctanh}(ax)}{x}$ & $\displaystyle \ln \! \left(a x \right) \mathrm{arctanh}\! \left(a x \right)-\frac{\mathrm{dilog}\! \left(a x +1\right)}{2}-\frac{\ln \! \left(a x \right) \ln \! \left(a x +1\right)}{2}-\frac{\mathrm{dilog}\! \left(a x \right)}{2}$ & $\displaystyle -\frac{\mathrm{dilog}\! \left(a x +1\right)}{2}+\frac{\mathrm{dilog}\! \left(-a x +1\right)}{2}$ \\ 

\hline

$\displaystyle \frac{2 \cos \! \left(3 x \right) \sin \! \left(3 x \right) {\mathrm e}^{-9 x}}{3}$ & $\displaystyle -\frac{2 \,{\mathrm e}^{-9 x} \cos \! \left(6 x \right)}{117}-\frac{{\mathrm e}^{-9 x} \sin \! \left(6 x \right)}{39}+\frac{2 \,{\mathrm e}^{-9 x} \cos \! \left(2 x \right)}{85}+\frac{9 \,{\mathrm e}^{-9 x} \sin \! \left(2 x \right)}{85}+\frac{{\mathrm e}^{-9 x} \left(-9 \sin \! \left(2 x \right)-2 \cos \! \left(2 x \right)\right)}{85}$ & $\displaystyle -\frac{{\mathrm e}^{-9 x} \left(2 \cos \! \left(6 x \right)+3 \sin \! \left(6 x \right)\right)}{117}$ \\

\hline

$\displaystyle \frac{\mathrm{i} \cos \! \left(x \right) -\sin \! \left(x \right)}{\mathrm{i} \sin \! \left(x \right) +\cos \! \left(x \right)}$ & $\displaystyle \mathrm{i} \ln \! \left(\sin \! \left(x \right) +\cos \! \left(x \right)\right)$ & $\displaystyle \mathrm{i}x$ \\ 

\hline

$\displaystyle \frac{1}{\left(x^{3}+2\right)^{3}}$ & $\displaystyle \frac{x}{12 \left(x^{3}+2\right)^{2}}+\frac{5 x}{72 x^{3}+144}+\frac{5 \,2^{\frac{1}{3}} \ln \! \left(x +2^{\frac{1}{3}}\right)}{216}-\frac{5 \,2^{\frac{1}{3}} \ln \! \left(x^{2}-2^{\frac{1}{3}} x +2^{\frac{2}{3}}\right)}{432}+\frac{5 \,2^{\frac{1}{3}} \sqrt{3}\, \arctan \! \left(\frac{\sqrt{3}\, \left(2^{\frac{2}{3}} x -1\right)}{3}\right)}{216}
$ & $\displaystyle \frac{\frac{5}{72} x^{4}+\frac{2}{9} x}{\left(x^{3}+2\right)^{2}}+\frac{5  \left( \sum\limits_{R=RootOf(\_z^3+2} \frac{\ln(x - \_R)}{\_R^2} \right)}{108} $ \\

\hline

$\displaystyle \frac{{\mathrm e}^{-\frac{\ln \left(x -1\right)}{2}-\frac{\ln \left(x +1\right)}{2}} x a}{x^{2}-1}$ & $\displaystyle -{\mathrm e}^{-\frac{\ln \left(x -1\right)}{2}-\frac{\ln \left(x +1\right)}{2}} a$ & $\displaystyle -\frac{a}{\sqrt{x -1}\, \sqrt{x +1}}$ \\

\hline

\end{tabular}
\end{table}

\subsection{Is the Classification Stage Necessary?}
\label{sec:results/2stage_analysis}

\begin{figure}[htbp]
  \centering
  \begin{subfigure}[b]{0.49\textwidth}
    \includegraphics[width=\textwidth]{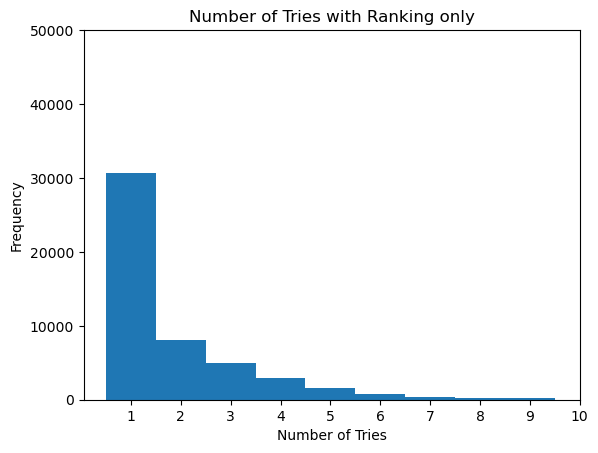}
    \caption{Attempts with just ranking}
    \label{fig:num_tries_rank}
  \end{subfigure}
  \hfill
  \begin{subfigure}[b]{0.49\textwidth}
    \includegraphics[width=\textwidth]{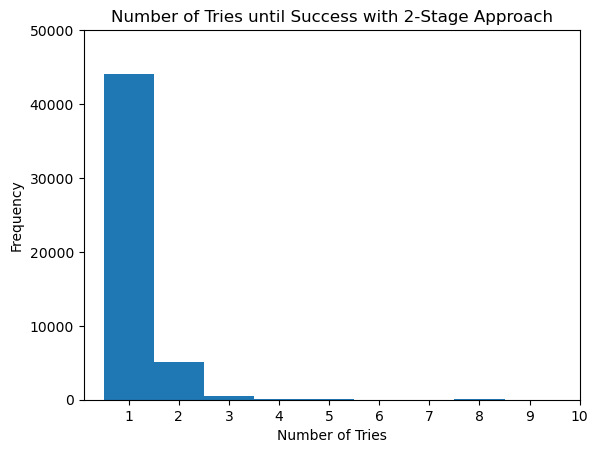}
    \caption{Attempts with classification + ranking}
    \label{fig:num_tries_2stage}
  \end{subfigure}
  \caption{The number of attempts the model takes to produce an answer. Adding a classification stage first (right) reduces the overall number of attempts compared to only ranking (left).}
  \label{fig:num_tries}
  \Description{Number of tries}
\end{figure}

As mentioned in Section~\ref{sec:methodology/ranking}, the classification stage is primarily to save compute time. Recall that the classification stage in our two-stage architecture only predicts if a method will produce an answer or not; not whether it gives the \textit{best} answer. The reason we include this stage is that on some expressions (especially non-elementary ones), a method can take a long time to finish. Furthermore, there is no guarantee that such a method will even produce an answer, meaning we wasted time on a method that does not work. To see the effects of the two-stage framework compared to just the ranking model alone, we counted the number of attempts (methods tried in the ranking) that it takes to produce an answer, as visualised in Figure~\ref{fig:num_tries}. This evaluation was on the 50,000 elementary examples from the hold-out test set. We do not include the special functions here because they have on average less applicable methods to use.  

Figure~\ref{fig:num_tries} shows the importance of the classification stage: the two-stage framework tries significantly fewer methods before finding an answer than ranking alone.  The two-stage framework can predict an applicable method on its first try $43.8\%$ more often than ranking only. The two-stage framework takes $57,411$ attempts over the entire 50,000 example dataset, compared to $93,845$ if just ranking. This means that the ranking only framework tries a staggering $63.5\%$ more method calls. 

However, there is a trade-off in including classification in the two-stage framework. It is possible for the ML model to mistakenly drop the best method during the classification stage, which then leads us to select a sub-optimal answer. This means that there is a trade-off between the number of attempts and accuracy. To illustrate this, consider the following hypothetical example in Table~\ref{tab:2stage_example}.  If there were only a ranking stage then we would try method 3 first which produces FAIL, and then move to method 1 which succeeds and is the best method for this example. However, if there were a two-stage framework, methods 1 and 3 did not pass and so we try method 2 which succeed, but it was not the optimal method. The ranking alone took two attempts but got the best answer, whereas the two-stage framework only took one attempt, but produced a sub-optimal answer. This is a simply hypothetical, but such examples do exist in the dataset. We find that we end up dropping just under 2\% in accuracy of exact matches, from 89.6\% when ranking alone to 87.8\% in the two-stage framework. We find that this relatively small drop in accuracy is acceptable, given that we try $63.5\%$ fewer methods overall, but the decision on that trade-off might go another way for a different application.

\begin{table}[ht]
  \centering
  \small
  \setcellgapes{3pt}\makegapedcells
  \caption{An example where a two‐stage framework can produce a sub‐optimal answer by including the classification stage.}
  \label{tab:2stage_example}
  \begin{tabular}{@{} l | c c c @{}}
    \toprule
    \thead{Method}
      & \thead{True Rank}
      & \thead{Classification Prediction\\(Stage 1)}
      & \thead{Rank Prediction\\(Stage 2)} \\
    \midrule
    Method 1 & 1    & \xmark & 2 \\
    Method 2 & 2    & \cmark & 3 \\
    Method 3 & FAIL & \xmark & 1 \\
    \bottomrule
  \end{tabular}
\end{table}


\section{Conclusions}
\label{sec:conclusion}

We developed ML models that are able to accurately rank methods for symbolic indefinite integration, improving the output quality of the Maple CAS from the current state-of-the-art. 

Perhaps the most notable finding is in data representation: we can clearly conclude that ML models which represent mathematical expressions as trees performed stronger than their sequential counterparts at this task.  We hypothesise this may also be useful for other AI tasks dealing with mathematical expressions.  We also found the framing of the problem very important: switching to a two-stage classification + ranking framework improved the results a lot compared to binary classification, and seems to be the sensible choice for a method selection problem such as this.

The Tree Transformer performed the best overall, getting close to 90\% on the test set.  More importantly, it showed strong generalisation abilities when evaluated on the Maple test suite, comfortably beating Maple's method selector on exact matches, with accuracy raising to 80.4\% if we allow answers within 10\% of the optimum.  These strong generalisation abilities required both the tree architecture and the framing of the ranking problem to unlock.  However, we should be clear that the models are still not able to generalise perfectly on OOD data, and more work is needed to understand what the remaining barriers are.

While we chose to focus on the DAG size as our metric, the implementation could easily have been adapted to focus on computation time instead (ranking from fastest to slowest). We hypothesise that our methodology could be used on similar problems of CAS optimisation. For example, the choice of different resultant algorithms was studied as a classification task in \cite{Simpson2016}, but ranking the methods as we do here should lead to further gains. In general, using ML in a ranking framework shows strong promise for any algorithm selection task where the goal is to return an ordered list of algorithms.

\subsection{Future Work}

As mentioned before, many ML symbolic mathematics experiments have made use LSTMs and transformers. However, other than our own preliminary work \cite{Barket2024_TreeLSTM}, we are aware of very limited cases where the tree structure is taken into account. This could have many applications to ML for other types of mathematical data, including automatic theorem proving \cite{Blaauwbroek2024_TheoremProving}, symbolic regression \cite{Kamienny2022_SR}, cylindrical algebraic decomposition \cite{Jia2023_CAD, Pickering2024_CAD}, Gr{\"o}bner bases \cite{kera2024_groebner}, and much more. 

One particular task we would like to focus on is a post-hoc analysis of symbolic integration in the original Lample \& Charton experiment. When the model is trained, are we able to see if the embeddings make sense, especially with this new tree positional encoding? Could we analyse the attention map to understand why certain operators or operands attend to one another?

There has also been further progress on modifications to ML architecture to better understand tree-structured data. The tree transformer used here relied on the positional encoding to explicitly represent the expression as a tree. Kogkalidis et al. \cite{Kogkalidis2024_algebraicPE} propose a new positional encoding that is based on group theory and is also suitable for trees. They are able to show that their encoding scheme outperforms the tree positional encoding that we have used here, and this would be a natural progression to try here in our experiment. One can also focus on modifying the attention mechanism rather than the positional encoding. Wang et al. \cite{Wang2019_TreeAttention} introduce constituent attention, where nearby nodes in the tree attend to each other more strongly in earlier transformer layers. In the context of mathematical expressions, this would mean forcing a node to attend more to its parent, sibling, and children, effectively focusing on sub-expressions within the overall expression. These are promising tree-based approaches to try, and there is an abundance of symbolic mathematics tasks to experiment with them on.

\subsection*{Data Access Statement}  The code and data underpinning the results in this paper are openly available here \url{https://zenodo.org/records/16780106}. However, this project will be updated as we make minor improvements for shipping into Maple. To get the most up-to-date version, visit the GitHub page: \url{https://github.com/rbarket/Ranking_Symbolic_Int_Methods/tree/main}. 

\begin{acks}
Matthew England is supported by
EPSRC Project EP/T015748/1, Pushing Back the Doubly-Exponential Wall of Cylindrical Algebraic Decomposition (the DEWCAD Project). Rashid Barket is supported by a scholarship provided by Maplesoft and Coventry University.    
\end{acks}

\bibliographystyle{plainurl}
\bibliography{ref.bib}

\appendix

\section{Generating Non-Elementary Functions}
\label{sec:special_functions}

Section~\ref{sec:data/generation} describes the data generation methods we used in order to produce (integrand, integral) pairs. These methods have been shown in our previous work to provide a good variety of elementary integrands \cite{Barket2023_generation}, \cite{Barket2024_Liouville}.  However, we have not before explored their use in generating non-elementary (or special) expressions. As we want the ML method selector to be as general as possible, this of course means training on non-elementary  functions as well. Here, we describe how to modify some of the data generators to produce non-elementary (integrand, integral) pairs. 

Note that non-elementary integrands are much harder to integrate. If a random expression is created and Maple is used to try to integrate the expression (the FWD data generation method), it will fail almost all of the time. We find that this is even true for the BWD method in regards to non-elementary functions: differentiating a relatively large non-elementary function and trying to integrate it with Maple still produces a failure in many cases. Of course, we know the integral this time, but we cannot label the data with Maple DAG sizes if Maple itself cannot integrate the expression. Thus, we must take careful care on how we create random non-elementary expressions.

Although there is no comprehensive list of non-elementary functions, Maple has a list of \textit{all} functions (elementary and non-elementary) on their help page: \underline{\url{https://www.maplesoft.com/support/help/Maple/view.aspx?path=initialfunctions}}. The natural first approach to generating a non-elementary function would be what Lample \& Charton \cite{Lample2020} did for generating random elementary expressions: construct a random unary-binary tree with empty nodes and then fill those nodes with the appropriate operators and operands. 

This is what we initially tried, but we failed to integrate many of these generated non-elementary expressions in both the FWD and BWD methods. There are two reasons for this: (1) an expression with even two or more special functions composed together is hard to integrate, and (2) there is a relationship between some special functions and having two or more unrelated special functions in the same expression makes it extremely hard to integrate (and also unlikely to be a problem we need to solve in practice). To address these concerns, we simplify the random expression generation and group relative special functions together when randomly sampling functions for expression generation. The groupings of special functions are shown in Table~\ref{tab:special_functions}, guided by the domain experts at Maplesoft. When we create a random expression, we only select from one of these groups. 

\begin{table}[ht]
    \centering
    \renewcommand{\arraystretch}{1.5}
    \caption{Groupings of special functions. Note that a special function can belong to more than one group.}
    \begin{tabular}{|c|p{9cm}|}
        \hline        
         \textbf{Group Name} & \textbf{Functions} \\
         \hline
         Bessel Functions & BesselI, BesselJ, BesselK, BesselY, AngerJ, WeberE, HanelH1, HankelH2, StruveH, StruveL \\
         \hline
         One parameter GAMMA Functions & GAMMA, lnGAMMA, Psi, Zeta, harmonic \\
         \hline
         Two parameter GAMMA Functions & GAMMA, Psi, Zeta, harmonic \\
         \hline
         Orthogonal Polynomials & ChebyshevU, ChebyshevT, HermiteH, LaguerreL, LegendreP, \newline LegendreQ \\
         \hline
         One parameter Elliptic Functions & EllipticE, EllipticCE, EllipticK, EllipticCK \\
         \hline
         Two parameter Elliptic Functions & EllipticE, EllipticF ,JacobiAM, JacobiCD, JacobiCS, JacobiCN, JacobiDC, JacobiDS, JacobiDN, JacobiNC,                          JacobiNS, JacobiND, JacobiSC, JacobiSN, JacobiSD \\
         \hline
         Error Functions & erf, erfc, erfi, dawson, FresnelC, FresnelS, Fresnelg, Fresnelf \\
         \hline
         Elliptic Integral Functions &  Ei, Si, Ci, Ssi, Shi, Chi \\
         \hline
         polylog & polylog \\
         \hline
         dilog & dilog \\
         \hline
         Piecewise Constant Functions & Re, Im, conjugate, abs, argument, signum, floor, ceil, round \\
         \hline
    \end{tabular}
    \label{tab:special_functions}
\end{table}

\subsection{FWD and BWD Generators}

The FWD and BWD methods are fairly straightforward: a random expression is created, and then it is integrated or differentiated. To create a random expression that contains special functions, the resulting expression must be much simpler than when creating purely elementary functions. Algorithm~\ref{alg:special_func} describes how to create a non-elementary expression, which then gets used in the FWD method or differentiated in the BWD method. In essence, we start with a base special function and apply a binary operator (e.g. $+,-,\times,\div$) to combine the base term and a randomly selected new term (elementary or non-elementary). In practice, we never set the number of terms higher than three. This does not guarantee that the FWD method or BWD method will work, but it is much more likely compared to randomly generating a unary-binary tree and filling in the nodes. 

\begin{algorithm}[ht]
\caption{Non-Elementary Function Creation}
\begin{algorithmic}[1]  
\Require A list of special functions $F$ 
\Statex \quad \quad A set of possible parameters $P$ that are given to a function in $F$
\Statex \quad \quad  Total terms \texttt{num\_terms}
\State Let \( f \in F \) be chosen uniformly at random.
\State Select a random subset \( \{p_1, p_2, \dots, p_k\} \subseteq P \), where \( 1 \leq k \leq n \), and pass as arguments to \( f \) to form a valid expression. 
\Comment{First term is always a special function}

\For{i from 1 to \texttt{num\_terms}}:
\State Let \(\texttt{is\_special} \in \{0, 1\} \) be chosen uniformly at random

\If{\texttt{is\_special}=1}
    \State Sample $g \in F$ and \( \{q_1, q_2, \dots, q_k\} \subseteq P \) to form a new special term
    \Else
        \State Create $g$ as a random elementary expression as in \cite{Lample2020}.   
\EndIf
\State Let \texttt{op} be randomly sampled from $\{+,-,\times,\div\}$ 
\Comment{Can be weighted}
\State $f \gets f \texttt{ op } g$ 
\EndFor

\State \Return \( f \)
\end{algorithmic}
\label{alg:special_func}
\end{algorithm}

\subsection{LIOUVILLE Generator}

Unlike the RISCH generator, the LIOUVILLE generator in Algorithm~\ref{alg:liouville_gen} is able to include special functions with only a little modification. This is because the LIOUVILLE generator is based on the parallel Risch algorithm, which is somewhat similar to the BWD method. To explain the modifications needed for the LIOUVILLE generator, we first explain how it works in as presented originally in \cite{Barket2024_Liouville}. The LIOUVILLE generator starts with in a field and adds $n$ field extensions, which can be exponential, logarithmic, tangent, or a special function. We have previously explored the first three cases, and now adapt the LIOUVILLE generator to work with special field extensions. The main difference is in how the field extensions $\theta_0,\dots,\theta_n$ are chosen. 

In the elementary cases, we randomly create each extension and then proceed with going through the steps of Algorithm~\ref{alg:liouville_gen}. This would not work with special functions because of the return steps in Lines 8 and 11. Here, a derivative of the created expression is made, like in the BWD method.  This creates issues because the derivative of a special function can introduce new special functions that were not defined in the input of the LIOUVILLE generator. This still means the expression is integrable, but we lose the benefits of the LIOUVILLE generator and this becomes the BWD method. For example, suppose $\theta_0=x \text{ and } \theta_1=\text{polylog}(x)$. The derivative of $\theta_1$ is $\Psi(x)$. If $\Psi(x)$ is not one of $\theta_0, \dots , \theta_n$, then the LIOUVILLE method is not guaranteed to work. 

Thus, the main change we made is in the input. First, a function $f$ is randomly chosen from Table~\ref{tab:special_functions}, and then it is differentiated to see what new extensions are needed. We get that $\theta_0=x, \theta_n=f$ and $\theta_1,\dots,\theta_{n-1}$ will be the functions that are produced from differentiating $f$.      

\begin{algorithm}[ht]
    \caption{Overview of Liouville Generator}
    \label{alg:liouville_gen}
    
    \begin{algorithmic}[1]
    \Require List of field extensions $T=[\theta_0,...,\theta_n]$ where $\theta_0=x$.
    \Statex \quad \quad \quad An upper bound on the multiplicity of the denominator $r$.
    \Statex \quad \quad \quad A boolean flag \texttt{normal} that determines whether to return the answer in normalised or partial fraction form.
    \Ensure $F', F$ such that $F'$ is an integrand and $F$ is its integral.
    
    \State Generate polynomials $q_1,...,q_r$ in $\theta_n$ with coefficients in $\theta_0,...,\theta_{n-1}$.
    \label{step:gen_qs}
    
    \State Let $D= \texttt{SquareFreeFactor}(q_1^1 \dots q_{r}^{r}) = Q_1^1 \dots Q_s^s$ where $s \geq r$. \Comment{Denominator}
    \label{step:gen_D}
    
    \State Generate polynomial $N$ in $\theta_n$ with coefficients in $\theta_0,...,\theta_{n-1}$ which has total degree smaller or equal to the total degree of $D$. \Comment{Numerator}
    \label{step:gen_N}
    
    \State Choose $j \leq s$. Generate $a_0,...,a_j \in \{Q_1,...,Q_s\}$ and $c_0,...,c_j \in \mathbb{F}$. Let $A~=~\sum_{i=0}^j c_i\log{(a_i)}$.
    \label{step:gen_as}
    
    \State Choose $k \in \mathbb{N}_0$. Generate $b_0,...,b_k \in \mathbb{F}[\theta_0,...,\theta_n]$ and $d_0,...,d_k \in \mathbb{F}$.  Let $B~=~\sum_{i=0}^k d_i\log{(b_i)}$.
    \label{step:gen_bs}

    \If{\texttt{normal}}
    \State $\hat{F} = \frac{N}{D}$;
    \label{step:gen_F_norm1}
    \State \textbf{return} $\texttt{Normalise}(\hat{F}' + A') + B',  \texttt{Normalise}(\hat{F} + A) + B$
    \label{step:gen_F_norm2}
    \label{step:gen_return_norm}
    \Else
    \State $G = \texttt{PartialFraction}(\frac{N}{D}) + A + B$
    \label{step:gen_F_PF}
    \State \textbf{return} $G', G$
    \EndIf
    
    \end{algorithmic} 
    
\end{algorithm}

\newpage

\section{Regression for Integration Method Selection}
\label{sec:regression}

In Section~\ref{sec:methodology}, we discussed how symbolic integration method selection can be framed in two different ways: classification or ranking. Before trying ranking, another framing we attempted was regression. 

\subsection{Motivation for Regression}

We were motivated by the substantial difference in DAG sizes that were possible. Consider the following two problem instances for the integration method \texttt{orering} we have in the dataset:

\begin{align}
\label{eqn:ex1}
\int x^{10} &= \frac{x^{11}}{11}  \\  
\label{eqn:ex2}
\int x^{10}-1 &= \frac{x(x^{10}-11)(x^{10}-1)}{11(x^4-x^3+x^2-x+1)(x^4+x^3+x^2+x+1)(x-1)(x+1)}  
\end{align}

Suppose the presented answer for the integrals in Equations (\ref{eqn:ex1}) and (\ref{eqn:ex2}) are the shortest answer possible, and so these instances are both labelled as 1 for binary classification in \texttt{orering}'s model. This does not take into account the different magnitude in their size: the DAG size in (\ref{eqn:ex1}) is much smaller than the DAG size in (\ref{eqn:ex2}) and thus we hypothesise this should be reflected in the loss score. This motivates us to look at predicting the DAG size rather than predicting if the method simply outputs the best answer, moving the paradigm from classification to regression. Then once the DAG size is predicted, we take the method with the lowest predicted DAG size as the method to use. A similar comparison of paradigms was made for the problem of selecting the variable ordering for cylindrical algebraic decomposition in \cite{Del2024_lessons}. 

\subsection{Imputation Difficulties}

One problem we uncovered in changing from classification to regression is \emph{imputation}. If the method fails for a problem, there is no DAG size to label the data with. This means we have to impute a value for the DAG size of a method. Imputing the minimum of all the DAG sizes results in the method being used more often than needed, while using the maximum has the opposite effect. We used the average in our experiments to try to address this. However, our models still do poorly on samples where the value was imputed. This gave us the original motivation to try a two-stage framework, similar to that we settled upon in Section \ref{sec:methodology/ranking}.

\subsection{Two-Stage Classification + Regression Approach}

\begin{figure}[htbp]
  \centering
  \begin{subfigure}[b]{0.7\textwidth}
    \includegraphics[width=\textwidth]{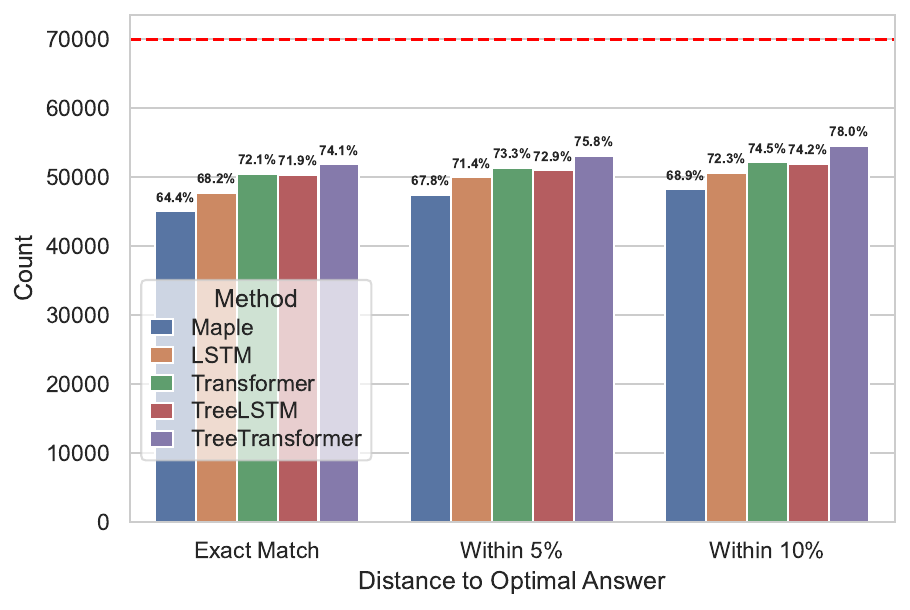}
    \caption{Results on the hold-out test set}
    \label{fig:results/regression/test}
  \end{subfigure}
  \hfill
  \begin{subfigure}[b]{0.7\textwidth}
    \includegraphics[width=\textwidth]{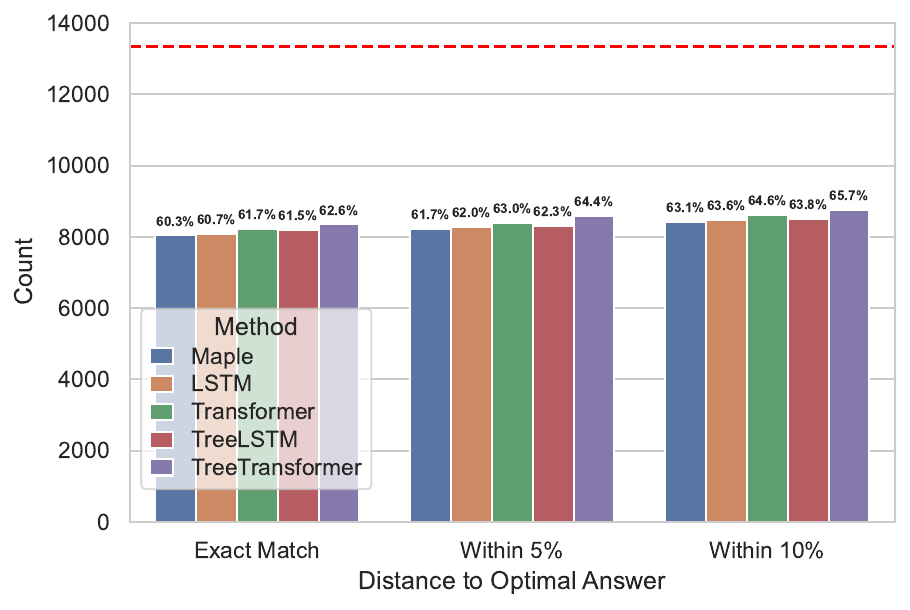}
    \caption{Results on the Maple Test Suite}
    \label{fig:results/regression/maple}
  \end{subfigure}
  \caption{Results on the two-stage classification + regression framework.}
  \label{fig:results/regression}
  \Description{Results on the hold-out test set and Maple test suite on the two-stage classification and regression framework. }
\end{figure}

We then tried a framework with a Stage 1 that trained an ML model to classify whether a given method will be \textit{successful} or not (as in Section~\ref{sec:methodology/ranking}) but then in Stage 2 we trained one ML regression model for each integration method to predict the DAG size. The shortest predicted regression from the methods that still passed Stage 1 is outputted. In the unlikely event where none of the ones predicted successful in Stage 1 work, we try all the methods in Stage 2.  The results of this two-stage classification + regression are shown in Figure~\ref{fig:results/regression}.

The results over the hold-out test set (Figure~\ref{fig:results/regression/test}) are as expected: the TreeLSTM and both transformers do better than Maple's method selector just as in the binary classification framework. This margin is now wider, showing that regression was an improvement, but but not as strong as the ranking model we settled upon finally. The exciting gains are seen in the external Maple test suite (Figure~\ref{fig:results/regression/maple}). Recall that in classification alone, the models were equal or only slightly better than Maple on exact matches. We see that this is not the case any more with the two-stage regression framework. Both the TreeLSTM and Transformer are now able to outperform Maple on this independent test set. 

\subsection{Motivation for Ranking}

Although regression is doing better than classification, there are still issues arising. The main reason we then switched from regression to ranking is that regression is \textbf{order invariant}. To understand this consider the hypothetical example of two ML models predicting DAG sizes for two different integration methods given in Table~\ref{tab:regression_example}. 

\begin{table}[ht]
  \centering
  \renewcommand{\arraystretch}{2}
  \caption{Model predictions vs.\ true scores for Method 1 and Method 2}
  \label{tab:regression_example}
  \begin{tabular}{lccll}
    \toprule
             & Method 1 & Method 2 & MSE & Order \\ 
    \midrule
    True scores   & 10       & 13       & —   & —     \\
    Model 1       & 14       & 13       & 
    \(\displaystyle \frac{1}{2}\bigl((10-14)^2 + (13-13)^2\bigr) = 8\) 
                 & \xmark \\
    Model 2       & 6        & 13       & 
    \(\displaystyle \frac{1}{2}\bigl((10-6)^2 + (13-13)^2\bigr) = 8\)  
                 & \cmark   \\
    \bottomrule
  \end{tabular}
\end{table}

The predictions of Model 1 and Model 2 both have a Mean Square Error (MSE) of 8, and they both predict the DAG size of Method 2 correctly. The key difference is their prediction of the DAG size for Method 1: Model 1 guesses above the true value (and higher than Method 2), whereas model 2 guesses below the true value. Although both models have the same MSE, it is only Model 2 that predicted the order correctly. This is what motivated us to switch from classification + regression to classification + ranking as presented in  Section~\ref{sec:methodology/ranking}.   

\end{document}